\documentclass[aps, prd,twocolumn,groupedaddress,showpacs,nofootinbib]{revtex4-1}
\pdfoutput=1

\usepackage{amssymb,amsfonts,amsmath,bm,dsfont}
\usepackage{graphicx}
\usepackage{hyperref}
\usepackage{multirow}
\usepackage[utf8]{inputenc}

\usepackage{amsmath}
\usepackage{mathtools}
\usepackage{amssymb}
\usepackage{units}
\usepackage{multirow}
\usepackage{slashed}
\usepackage{upgreek}
\usepackage{bbm}
\usepackage{color}
\usepackage{filecontents}
\usepackage{hepunits}

\newcommand{\FP}{Faddeev--Popov}

\newcommand{\DS}{Dyson--Schwinger}
\newcommand{\FRG}{Functional Renormalization Group}

\renewcommand{\vec}{\mathbf}

\newcommand{\hati}{\hat{\imath}}

\newcommand{\tr}{\mathrm{tr}}
\newcommand{\SU}[1]{\ensuremath{\mathrm{SU(#1)}}}
\newcommand{\groupU}[1]{\ensuremath{\mathrm{U(#1)}}}

\begin{document}

\date{\today}

\title{Gribov horizon and Gribov copies effect in lattice Coulomb gauge}
\author{Giuseppe~Burgio}
\email{giuseppe.burgio@uni-tuebingen.de}
\affiliation{Institut f\"ur Theoretische Physik, Auf der Morgenstelle 14,
 72076 T\"ubingen, Germany}

\author{Markus~Quandt}
\email{markus.quandt@uni-tuebingen.de}
\affiliation{Institut f\"ur Theoretische Physik, Auf der Morgenstelle 14, 
72076 T\"ubingen, Germany}
\author{Hugo~Reinhardt}
\email{hugo.reinhardt@uni-tuebingen.de}
\affiliation{Institut f\"ur Theoretische Physik, Auf der Morgenstelle 14, 
72076 T\"ubingen, Germany}

\author{Hannes~Vogt}
\email{hannes.vogt@uni-tuebingen.de}
\affiliation{Institut f\"ur Theoretische Physik, Auf der Morgenstelle 14, 
72076 T\"ubingen, Germany}

\begin{abstract}

Following a recent proposal by Cooper and Zwanziger \cite{Cooper:2015sza} we 
investigate via {$SU(2)$} lattice simulations the effect on the Coulomb gauge 
propagators and on the Gribov-Zwanziger confinement mechanism of selecting the 
Gribov copy with the smallest non-trivial eigenvalue of the Faddeev-Popov 
operator, i.e.~the one closest to the Gribov horizon. 
Although such choice of gauge drives the ghost propagator towards the prediction 
of continuum calculations, we find that it actually overshoots the goal. With
increasing computer time, we observe that Gribov copies with arbitrarily small 
eigenvalues can be found. For such a method to work one would therefore need 
further restrictions on the gauge condition to isolate the physically relevant 
copies, since e.g.~the Coulomb potential $V_C$ defined through the Faddeev-Popov 
operator becomes otherwise physically meaningless. Interestingly, the Coulomb 
potential alternatively defined through temporal link correlators is only 
marginally affected by the smallness of the eigenvalues.

\end{abstract}


\maketitle

\section{Introduction}

Coulomb gauge plays a prominent role in the Hamiltonian formulation of 
non-Abelian gauge theories 
\cite{Gribov:1977wm,Jackiw:1977ng,Christ:1980ku,Schutte:1985sd,Zwanziger:2002sh,Reinhardt:2011fq}; 
within this framework, 
variational \emph{Ans\"atze} offer a promising approach to determine the 
vacuum state \cite{Schutte:1985sd,Szczepaniak:2001rg,Feuchter:2004mk,Reinhardt:2004mm}. 
In the last years much effort has been invested in this 
direction, achieving a large number of interesting analytical results which 
combine to a rather concise picture of the low-energy sector in gauge 
theories, see e.g. Refs.~\cite{Zwanziger:2002sh,Schleifenbaum:2006bq,Reinhardt:2008ek}. 
{The picture of the vacuum conveyed by this approach is the Gribov-Zwanziger (GZ) confinement 
scenario, which in turn is based on a restriction of the functional integral to the 
first Gribov region. Applied to Coulomb gauge, this scenario leads to a number of 
general predictions which are not tied to the variational approach, and which can be 
accessed directly in lattice simulations:}
\begin{enumerate}
\item the Coulomb potential should be bound from below
\cite{Zwanziger:2002sh} by the physical Wilson potential \cite{Wilson:1974sk}, 
i.e.~the presence of Coulomb confinement should be a 
{\it necessary} condition for the physical confinement mechanism to take place;
\item the gluon dispersion relation 
should be infra-red (IR) divergent, naturally providing a confining scale 
\cite{Gribov:1977wm};
\item the Coulomb gauge ghost form factor should be
IR-divergent.
\end{enumerate}
{The variational approach of Refs.~\cite{Feuchter:2004mk,Reinhardt:2004mm,Schleifenbaum:2006bq,Epple:2006hv} 
realizes this scenario, provided that the third 
condition (often called \emph{horizon condition} in this context) is 
implemented as a boundary condition.\footnote{The horizon condition selects 
among several possible solutions in the variational approach, while it 
comes out self-consistently in the renormalization group approach
\cite{Leder:2010ji}. Physically, this can be interpreted 
as a vanishing dielectric constant of the vacuum, i.e.~a manifestation of 
the dual Meissner effect \cite{Reinhardt:2008ek}.} Therefore, a}
lattice investigation of the above listed Coulomb gauge correlators 
represents a powerful tool to gain insight in the 
mechanism of quark confinement while offering a direct bridge to continuum 
setups; this program has been thoroughly carried out in
Refs.~\cite{Burgio:2008jr,Burgio:2009xp,Quandt:2010yq,Burgio:2012ph,Burgio:2012bk,Burgio:2013naa,Burgio:2015hsa}. 
While the gluon sector has been found to agree with
the continuum predictions, confirming the dynamical generation of a Gribov mass 
$M \approx 0.9\,\mathrm{GeV}$ and the validity of 
Gribov's formula for the gluon propagator \cite{Burgio:2008jr,Burgio:2009xp}, 
the ghost sector was shown to agree only qualitatively with the continuum 
predictions. In particular, the IR divergence of the ghost form factor 
determined in lattice simulations \cite{Burgio:2012bk,Burgio:2013naa,Burgio:2015hsa}
is much weaker than the one 
predicted by continuum calculations \cite{Feuchter:2004mk,Schleifenbaum:2006bq,Epple:2006hv}, 
and a Coulomb string tension could 
be extracted from the IR behavior of the Coulomb potential only under very 
optimistic assumptions \cite{Burgio:2012bk,Burgio:2013naa,Burgio:2015hsa}.
Furthermore the lattice results are in conflict with the sum rule for the 
infrared exponents \cite{Schleifenbaum:2006bq}, which merely assumes that the ghost-gluon
vertex in Coulomb gauge is bare, or at least non-singular, in the deep infra-red.

In a recent work Cooper and Zwanziger \cite{Cooper:2015sza} have 
proposed to implement Coulomb gauge by picking the Gribov copy with the
lowest eigenvalue of the \FP\ operator, instead of the ``best copy'' (\emph{bc}) 
with the maximal value of the Coulomb gauge functional. They argue that a lattice 
simulation based on such a setup would lead to a better agreement with continuum 
predictions. 
The aim of this paper is to directly implement this proposal on the lattice 
and analyze its consequences on the correlators which should bear the 
signature of the Gribov-Zwanziger confinement mechanism.
As a by-product, we will also be able to re-analyze the \emph{bc} strategy with
very high statistics, as finding a small eigenvalue of the \FP\ operator 
requires the analysis of a very large number of gauge copies.

\section{The Gribov problem}

As Gribov has shown long ago \cite{Gribov:1977wm} the Coulomb gauge 
condition $\partial_i A_i = 0$, among others, is not sufficient to 
select a single configuration from the gauge orbit uniquely. 
On the lattice, gauge fixing amounts to selecting, for each given 
configuration $\{ U_\mu(x) \}$ of links, a gauge rotation 
$g(x) \in SU(N_c)$ such that some (unique) condition is met.
In particular, Coulomb gauge fixing is achieved by maximizing, 
for each time slice $t$, the functional
\begin{equation}
\label{eq:gribov:gff}
F_t^U[g] = \frac{1}{N_c N_d V}\sum_{\vec x,i} \mathsf{Re}\, \tr 
\left[ U^g_i(t,\vec{x}) \right]\,,
\end{equation}
where $V$ is the spatial volume of the lattice and the sum extends over all 
spatial links only. A \emph{local} maximum of \eqref{eq:gribov:gff} picks 
out - more or less randomly -- one copy in the first Gribov region (where the 
\FP\ operator is positive definite), out of many others that all satisfy the 
same condition. A unique prescription, which would solve the Gribov problem 
completely, would amount to finding the \emph{global} maximum, i.e.~the 
representative of the gauge orbit in the so-called fundamental modular region 
(FMR). Finding such a global maximum of a function with many degrees of freedom
is, however, analogous to finding the ground state of an \SU{N} spin glass 
\cite{Marinari:1991zv}, a problem which is known to be NP-hard even for the much 
simpler case of the $\mathbb Z_2$ gauge theory \cite{Barahona:1982}.

In the past, two approaches have been widely used to tackle the problem of Gribov 
copies in lattice gauge theory. The first one is to simply neglect that there 
is a problem at all, essentially stating that Gribov copies have no physical 
significance. In this case, the first (local) maximum found by the algorithm
is selected and one proceeds in calculating all relevant (gauge dependent) quantities.
In the literature, this process goes under the name of ``minimal gauge" 
\cite{Maas:2008ri}.\footnote{In the literature the term {\it minimal gauge} had originally 
been applied in Landau gauge to the representative of the fundamental modular region 
along the gauge orbit \cite{Cucchieri:1997dx,Cucchieri:1997ns}. Later the term 
{\it absolute gauge} stuck for this case, while minimal gauge was ``downgraded" to its 
present use \cite{Maas:2008ri}.}

The second approach is to choose the copy with the highest value of the gauge 
functional as the ``best representative'' of the global maximum, based on the 
conjecture that results for gauge dependent quantities will be
strongly correlated with the value of the gauge functional. 
In order to clarify this statement, let $\{U_\text{FMR}\}$ be the ensemble of 
gauge configurations which are in the FMR, i.e.\ $F[U_\text{FMR}] = \text{max}$,
and let $\{U_\text{bc}\}$ be the 
ensemble with gauge configurations close to such a maximum
\begin{equation}
 F[U_\text{FMR}] \gtrsim F[U_\text{bc}],
\end{equation}
i.e.\ the set of configurations which correspond to the best maximum one could 
find numerically. The assumption is that the $U_\text{bc}$ are, in some sense,
``close'' to the $U_\text{FMR}$, and this carries over to the expectation value
of any gauge variant quantity $\Omega$, i.e.
\begin{equation}
 \left< \Omega(U_\text{bc}) \right> \approx 
 \left< \Omega(U_\text{FMR}) \right> \equiv \langle \Omega \rangle_\text{phys}\,.
 \label{assump-1}
\end{equation}
No mathematical proof of this assumption exists, and a direct numerical test
is only feasable for toy models on very small lattices. One such test, a
\groupU{1} lattice theory on a 2-dimensional sphere, actually provides 
numerical evidence \emph{against} the hypothesis in Eq.~\eqref{assump-1}
\cite{deForcrand:1994mz}.
For historical reasons, we will call the ensemble $\{U_\text{bc}\}$ 
the \emph{best copy} (bc) ensemble. 

A third approach for resolving the Gribov problem has been discussed 
for Landau gauge in Refs.~\cite{Sternbeck:2012mf,Sternbeck:2013zja}: 
instead of choosing 
the copy with the best value of the gauge functional, one picks the copy for which 
the first non-trivial eigenvalue of 
the \FP\ operator is smallest, the so-called \emph{lowest copy} (lc). 
We will borrow this notation from the 
aforementioned papers. The idea behind the lc-approach is that this 
should choose configurations that are close to the Gribov horizon where the 
\FP\ operator becomes singular. According to Gribov's and Zwanziger's entropic 
reasoning, such configurations should be the relevant ones in the thermodynamic 
limit. The authors of Refs.~\cite{Sternbeck:2012mf,Sternbeck:2013zja} found 
that both the ghost dressing function and -- to a much smaller extent -- 
the gluon propagator are enhanced in the IR for the lowest-eigenvalue 
copy when compared to the bc-approach, while they become flatter if one 
chooses a copy with a large eigenvalue of the \FP\ operator instead. 
Similar attempts to tweak the Landau gauge fixing procedure in order to make the 
IR-behaviour of the ghost propagator match the decoupling solutions found in the 
continuum (eq.~by \DS\ or \FRG\ techniques) had previously been put forward 
with mixed results \cite{Maas:2009se,Maas:2015nva}.

As discussed in the introduction, a quantitative 
discrepancy exists in Coulomb gauge between the IR exponent of the ghost 
dressing function in the Hamiltonian variational approach 
\cite{Feuchter:2004mk,Reinhardt:2004mm,Schleifenbaum:2006bq,Epple:2006hv}
and the corresponding lattice results \cite{Burgio:2012bk,Burgio:2013naa,Burgio:2015hsa}. 
On the other hand the behavior of the gluon propagator agrees very well 
between the two approaches \cite{Burgio:2008jr,Burgio:2009xp}. 
Since the IR exponents of the ghost form factor and the gluon propagator should 
be related by a sum rule which is based on the sole assumption that 
the ghost-gluon vertex should be bare, or at least non-singular, in 
the deep infra-red \cite{Schleifenbaum:2006bq} (a fact that it 
known to hold in Landau gauge and expected to carry over to 
Coulomb gauge\footnote{In Landau gauge, the vertex is expected 
to be unrenormalized  based on Slavnov-Taylor identities \cite{Taylor:1971ff};
this is confirmed by lattice simulations which find only mild deviations from 
a bare vertex over the entire momentum range \cite{Cucchieri:2008qm}. A 
similar conclusion can also be made in Coulomb gauge within the variational approach
(using the continuum propagators as input) \cite{Huber:2014isa}.}),
this poses an unresolved puzzle. 

{One possible explanation for such disagreement is 
that the variational approach would have to be improved in order to
better reproduce the lattice results. This goes beyond a mere improvement
of the variational \emph{ansatz}, since the sum rule must hold for any ansatz 
(assuming a non-singular ghost-gluon vertex in the IR). One possible idea is that
the proper implementation of the GZ idea would go beyond the standard Coulomb
Hamiltonian combined with the horizon condition, and additional terms in the 
action or Hamiltonian  would be required, which could eventually reconcile 
the sum rule with lattice propagators. There are some indications that such 
a refinement is necessary in Landau gauge, where additional condensates can
be introduced in the GZ action in order to make the GZ scenario agree with 
lattice data \cite{Cucchieri:2011ig}. In the Hamiltonian approach, however,
we see no compelling evidence for such a modification, in particular since 
the present investigation will show that there is no such thing as 
``the lattice propagators'' in Coulomb gauge, at least with current 
computational power. It would then be very hard to identify the proper 
extension of the Coulomb gauge GZ scenario  required to match the 
inconsistent lattice data.}

{This leaves us with the second logical explanation for the sum-rule puzzle,
namely that the current lattice simulations in Coulomb gauge do not describe 
continuum physics and hence need refinement. More precisely,}
the bc-strategy on the lattice could be biased by artifacts 
related to the Gribov problem, being unable to come close enough to
the Gribov horizon, and the lc-strategy might provide a better 
description of continuum physics \cite{Cooper:2015sza}. 
To check this conjecture we will adapt in the following the 
lc-strategy to Coulomb gauge.

\section{Lattice Setup}

For our study {we use the colour group $G=SU(2)$ for simplicitly and employ}
the isotropic and the anisotropic Wilson gauge action \cite{Burgio:2003in}
\begin{align}
S = \sum_x \biggl\{&\beta_s \sum_{j> i=1}^3\left(1-
\frac{1}{2} \mathsf{Re}\,\tr{U_{ij}(x)}\right)\nonumber\\
+&\beta_t 
\sum_{i=1}^3
\left(1-\frac{1}{2}\mathsf{Re}\,\tr{U_{i4}(x)}\right)
\biggr\}\,,
\label{s_anis}
\end{align}
where we parameterize $\beta_s = \beta \gamma$ and $\beta_t = \beta/\gamma$, 
with $\gamma$ the bare anisotropy, while $\xi =  a_s/a_t$ denotes the 
renormalized anisotropy in the following. We have used isotropic lattices of 
three different sizes and discretizations in our analysis. Since the ghost 
propagator is known to suffer from strong scaling violations on isotropic 
lattices we include two anisotropic lattices of fixed size. Our setup is 
summarized in Tab.~\ref{tab:gribov:configs}. To fix the lattice spacing we 
used the {$SU(2)$} results known from the literature as summarized in the
tables given in Ref.~\cite{Burgio:2008jr}. We have also fixed 
$\sqrt{\sigma} = 0.44\, \giga \electronvolt$ to set the physical scale.

\renewcommand{\arraystretch}{1.4} 
\begin{table}[htb]
\centering
\begin{tabular}{cccccc}
\toprule
Label 	& \qquad Size\qquad\qquad  		& \quad$\xi$\qquad & \quad$\beta$ \quad\quad 	
& $a_s$ [$\giga \electronvolt^{-1}$] & $L$ [$\femto\meter$]\\ \colrule
A1	& $16^4$		& 1 	& 2.2 		& 1.07 		& 3.4	\\
A2	& $16^4$		& 1 	& 2.3 		& 0.84 		& 2.6	\\ 
A3	& $16^4$		& 1 	& 2.4 		& 0.61 		& 1.9	\\ 
B1	& $24^4$		& 1 	& 2.2 		& 1.07 		& 5.0	\\
B2	& $24^4$		& 1 	& 2.3 		& 0.84 		& 4.0	\\ 
B3	& $24^4$		& 1 	& 2.4 		& 0.61 		& 2.9	\\
C1	& $32^4$		& 1 	& 2.2 		& 1.07 		& 6.7	\\
C2	& $32^4$		& 1 	& 2.3 		& 0.84 		& 5.3	\\
C3	& $32^4$		& 1 	& 2.4 		& 0.61		& 3.8	\\ 
D1	& $128\times 32^3$	& 4 	& 2.25 		& 1.11 		& 7.0	\\ \botrule
 \end{tabular}
\caption{Lattice setup.}
\label{tab:gribov:configs}
\end{table}
\renewcommand{\arraystretch}{1.0}

\section{Gauge fixing and Gribov copies.}
\label{sec:gribov:gaugefixing}

\begin{figure}[phtb]
	\center
	\includegraphics[width=.494\columnwidth]{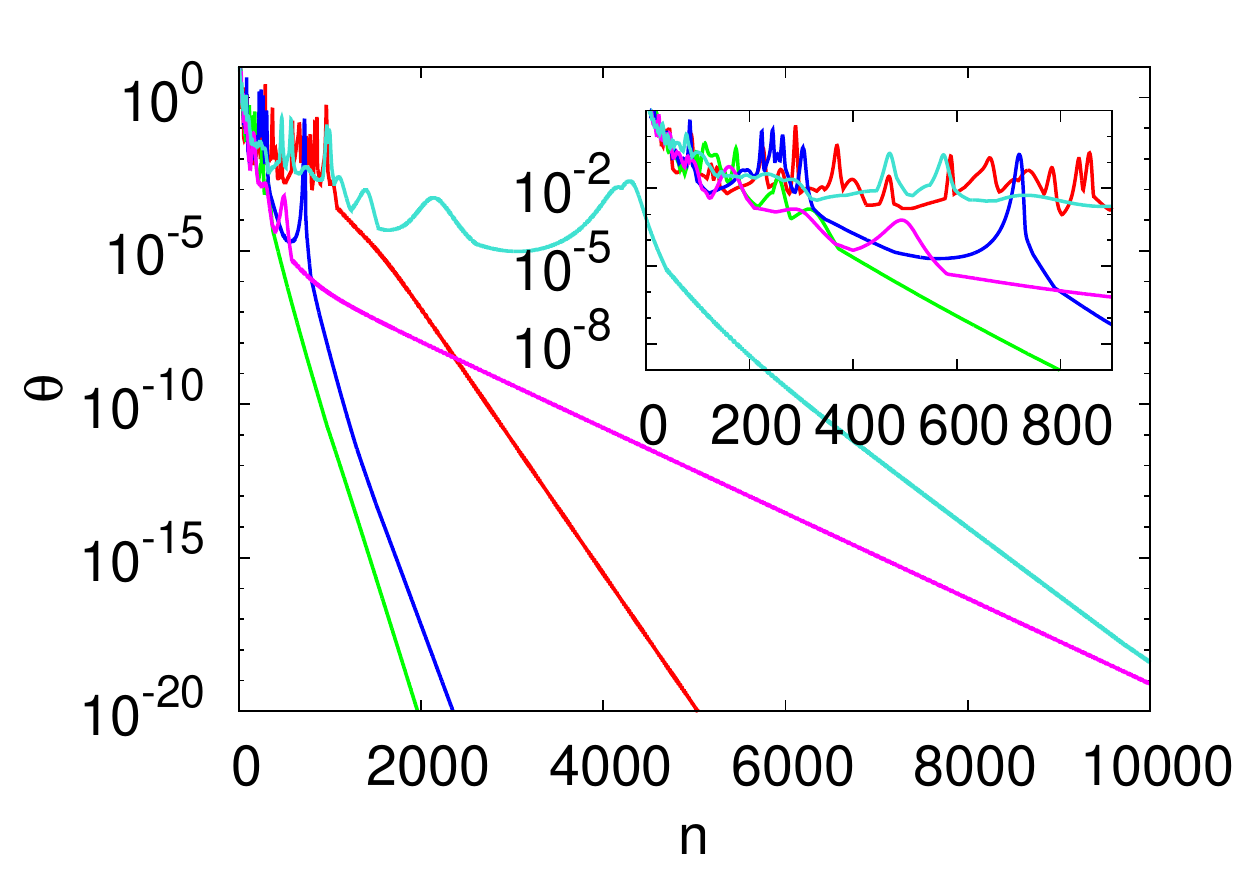}
	\includegraphics[width=.494\columnwidth]{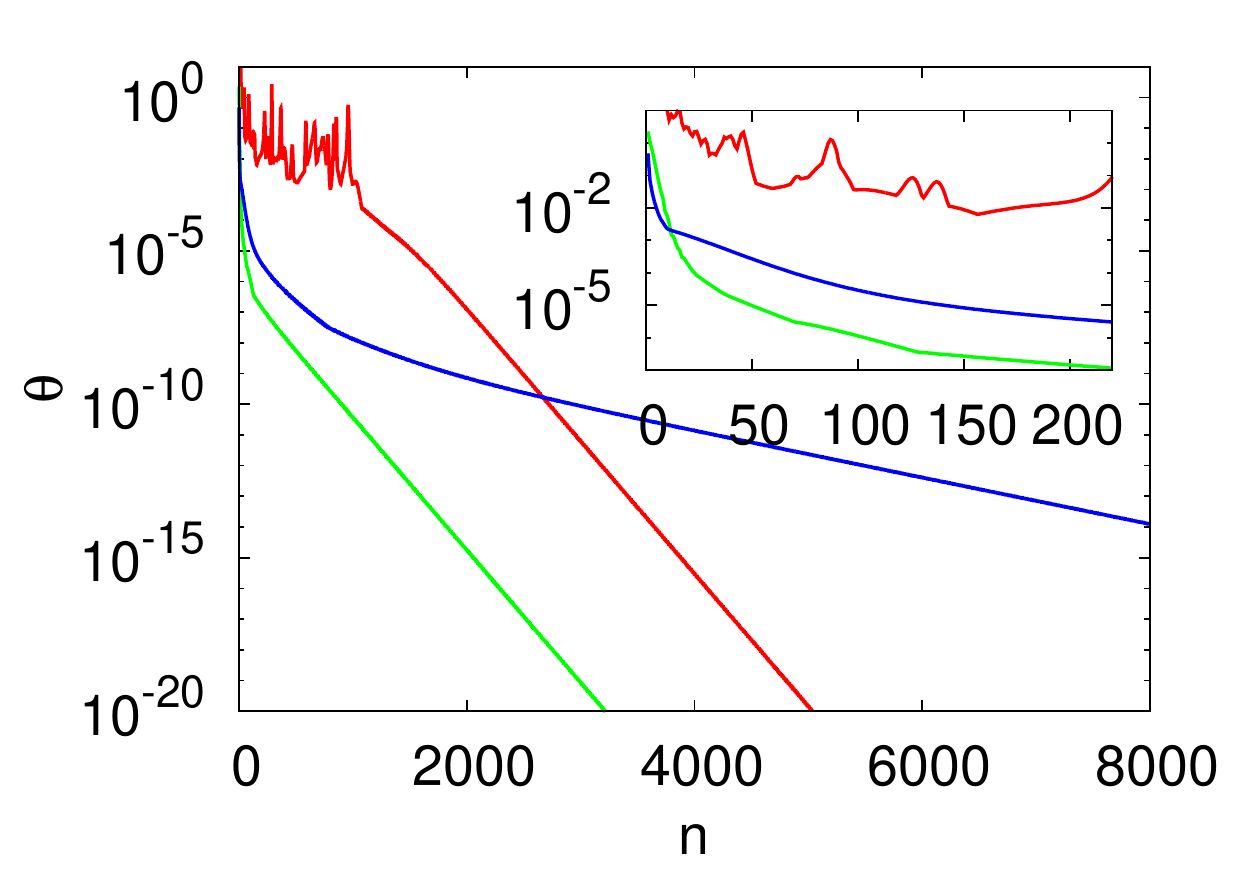}
	\caption{Gauge precision $\theta$ over the number of over-relaxation 
                 steps (color online). \emph{Left panel:}: 5 gauge copies of the same 
                 configuration. The light blue (top) curve is with simple relaxation, 
                 the other lines correspond to over-relaxation ($\omega=1.7$).
                 The pink line with the smallest slope corresponds 
                 to a significantly smaller value (compared to the other copies) of 
                 the first non-trivial eigenvalue $\lambda_1$ of the FP operator. 
                 \emph{Right panel:} the red (top) and green (bottom) 
                 line correspond to over-relaxation ($\omega=1.7$) without (red) and 
                 with (green) simulated annealing preconditioning. As can be seen,
                 the preconditioning removes the first phase where the algorithm 
                 tries to locate a maximum, while the slope of the second phase 
                 (the eventual convergence speed) is unchanged. The blue curve 
                 in the middle employs proconditioning with simple relaxation
                 ($\omega = 1$) and shows no fluctuating first phase but a 
                 much smaller convergence speed. All three lines converge towards 
                 the same Gribov copy, as was confirmed by identical functional 
                 values and an identical first non-trivial FP eigenvalue.}
\label{fig:gribov:oriter_example}
\end{figure}

Both for the lc and the bc strategy we use the over-relaxation 
technique \cite{Mandula:1990vs} in the CUDA implementation cuLGT 
\cite{Schroeck2013a}. In Ref.~\cite{Schroeck2013a}, simulated annealing 
\cite{Kirkpatrick:1983zz,Kirkpatrick1984} is also discussed as a technique to 
increase the probability to find the absolute maximum of the gauge fixing 
functional, i.e.\ to find a better best-functional copy. 
By now, the de-facto standard technique to find the (best approximation of the) global maximum
is a combination of repeated gauge fixing and a pre-conditioning with simulated 
annealing \cite{Bogolubsky:2005wf,Bogolubsky:2007bw}. In this context, repeated gauge 
fixing means to start the gauge fixing multiple times from a random gauge transformation 
and select the copy which best satisfies the bc (or lc) condition. 
In Fig.~\ref{fig:gribov:oriter_example} we show an illustrative plot of the evolution 
of the gauge fixing precision 
\begin{align}
\label{eq:landau:gaugeprec:max}
\theta \equiv \frac{1}{N_c} \max_{\vec{x}} \tr\left[ 
 \Delta(x)\Delta^\dag(x)\right]
\end{align}
with 
\begin{align}
  \Delta(x) \equiv \left[\partial_i A_i\right]^\text{lat} = 
\sum_i\left[A^\text{lat}_i(x)-A^\text{lat}_i(x-\hati)\right]
 \nonumber
\end{align}
over the number of gauge fixing steps. 
In the figure on the left hand side four runs with 
over-relaxation parameter $\omega = 1.7$ and one run with $\omega = 1$ (pure 
relaxation) are shown. The gauge fixing has two characteristic stages: in the 
first stage the precision is fluctuating strongly at a rather high value until 
a maximum is located with a precision of about $\theta \approx 10^{-4}$. Then, in the second 
stage, the precision monotonically approaches zero. As shown on the right hand 
side, if simulated annealing pre-conditioning is used the first 
stage is already overcome in the simulated annealing phase (which is not shown 
in the plot). 

As we focus on the lc-approach in this study, our goal is not to bias our 
algorithm towards copies with a high value of the g.f.~functional, and we 
thus have to waive simulated annealing preconditioning. Since the bc results 
in this chapter are mostly obtained as a byproduct of the main search for the 
lowest-eigenvalue copy, they are also not preconditioned with simulated annealing, 
unless explicitly stated otherwise. 
Unfortunately, no algorithm is known that would precondition the gauge fixing 
to a low eigenvalue of the \FP\ operator and we have to rely on pure 
over-relaxation with a high number of gauge copies $N_r$.

\begin{figure}[phtb]
\center
  \includegraphics[width=0.78\columnwidth]{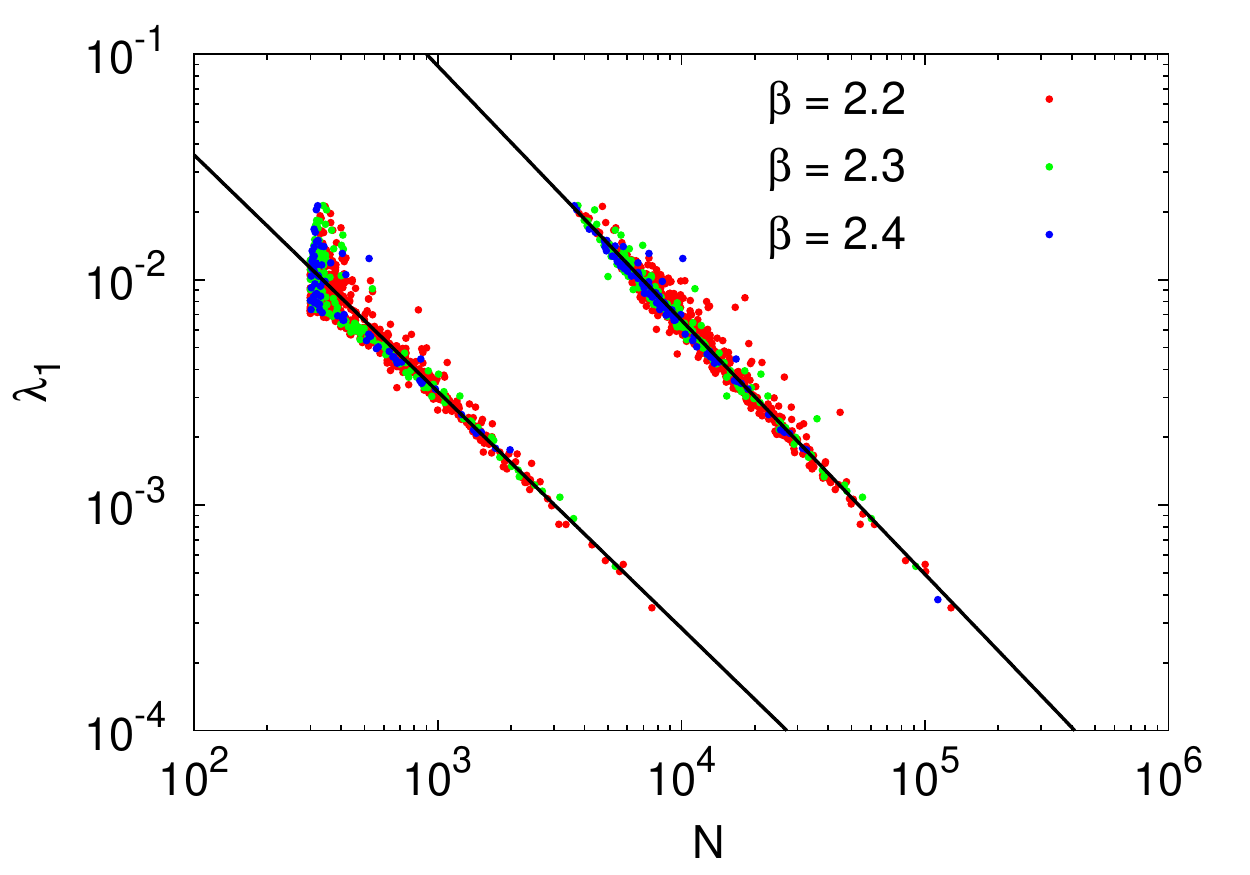}
 \caption{Smallest eigenvalue $\lambda_1$ of the \FP\ operator as a function of 
          the number of gauge-fixing iterations. From each set A1, A2 and A3, we used 10 
          configurations and calculated 1000 gauge-copies. The data points 
          which correspond to fewer iterations (left) are from runs with 
          $\omega = 1.9$, for the points with more iterations (right) we used 
          $\omega=1$.}
 \label{fig:gribov:iter_lambda_corr}
\end{figure}

In a first run we calculated the lowest eigenvalue $\lambda_1$ on
$N_r = \mathcal O (10^3)$ copies of the small lattices. In
Ref.~\cite{Greensite:2004ur} it was noticed that the size of the 
smallest eigenvalue is correlated 
with the number of gauge fixing iterations~$N_\text{it}$ that are necessary to achieve a 
given accuracy $\theta$, as indicated in 
Fig.~\ref{fig:gribov:oriter_example}. The 
reason for this behavior is that a low eigenvalue means an almost flat 
direction in the g.f.~functional and an ill-conditioned \FP\ operator, leading to a 
slow convergence of the iteration process. In 
Fig.~\ref{fig:gribov:iter_lambda_corr} we investigated this behavior in more 
detail. We find a perfect correlation of $\lambda_1$ and $N_\text{it}$, 
independently of the coupling $\beta$, with the slope only depending
very weakly on the over-relaxation parameter $\omega$. In fact, we find that 
all data can be perfectly described by the simple power law 
\begin{equation}
 \lambda_1\left( N_\text{it} \right) = \frac{c}{N_\text{it}^\gamma}.
\label{eq:eigen_copy}
\end{equation}
with $\gamma \approx 1.1$ and
the proportionality factor $c$ strongly depending on $\omega$.
To rule out that the over-relaxation parameter $\omega$ conditions the algorithm 
to find a gauge copy with specific value of $\lambda_1$, we verified that 
$\omega$ has no influence, on average, on how often a configuration with small 
eigenvalue is found. This is also indicated in 
Fig.~\ref{fig:gribov:iter_lambda_corr}, though in the plot it is obfuscated by 
the bulk around the minimal number of iterations.

The correlation of the number of iterations and the smallness of the \FP\ 
eigenvalue allows us to tweak our algorithm: since the calculation of 
eigenvalues is computationally the most demanding part in our gauge fixing 
program, we implemented a (self-adjusting) threshold, where the eigenvalues 
are calculated only for ``promising'' gauge copies for which the number of 
iterations exceeds a certain value. Since the smallest eigenvalue (which we 
are able to find) differs from configuration to configuration, 
we usually re-start with a small threshold for each configuration. 
If we do find a small eigenvalue, the threshold is updated to a factor $\alpha$ 
of the number of iterations that were needed to find this particular (small) 
eigenvalue. 
We find that $\alpha = 0.8$ provides a suitable update strategy: 
with this setting eigenvalues are calculated in a typical run for 
many gauge copies up to a point where a small eigenvalue is found and the 
threshold is changed. Since usually the Gribov copies with the smallest 
eigenvalue are well separated from the one with the next-to-smallest 
eigenvalue, this procedure constrains the program to only evaluate the 
eigenvalues for promising configurations with the smallest $\lambda_1$.

\begin{figure}[phtb]
  \includegraphics[width=0.98\columnwidth]{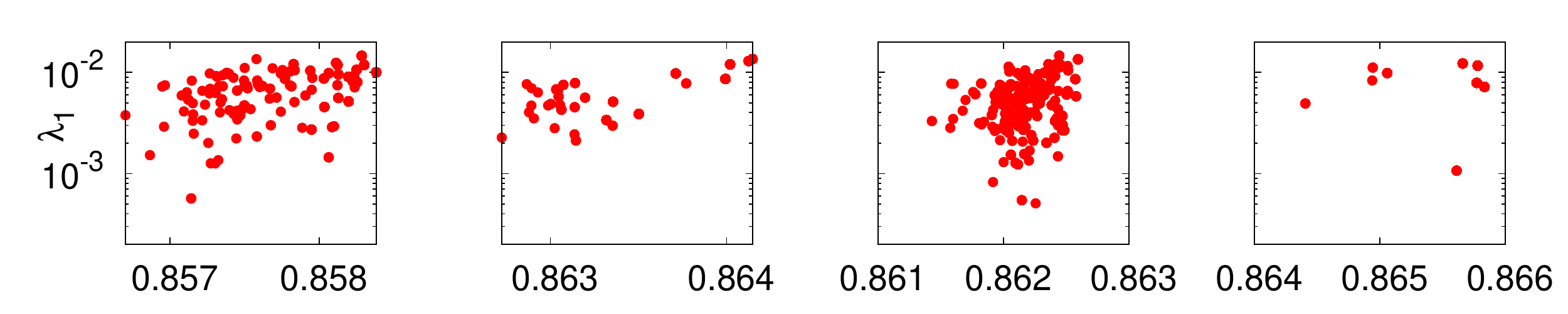}
  \includegraphics[width=0.98\columnwidth]{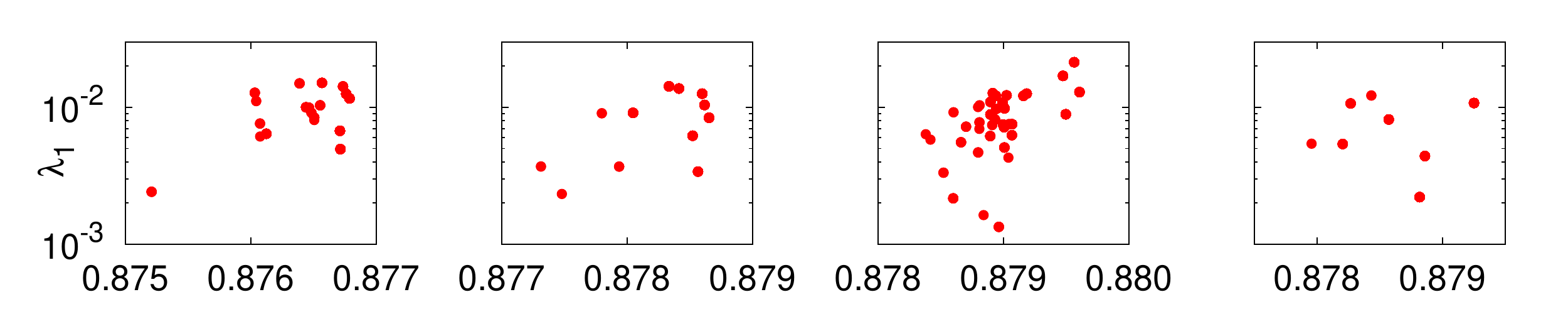}
  \includegraphics[width=0.98\columnwidth]{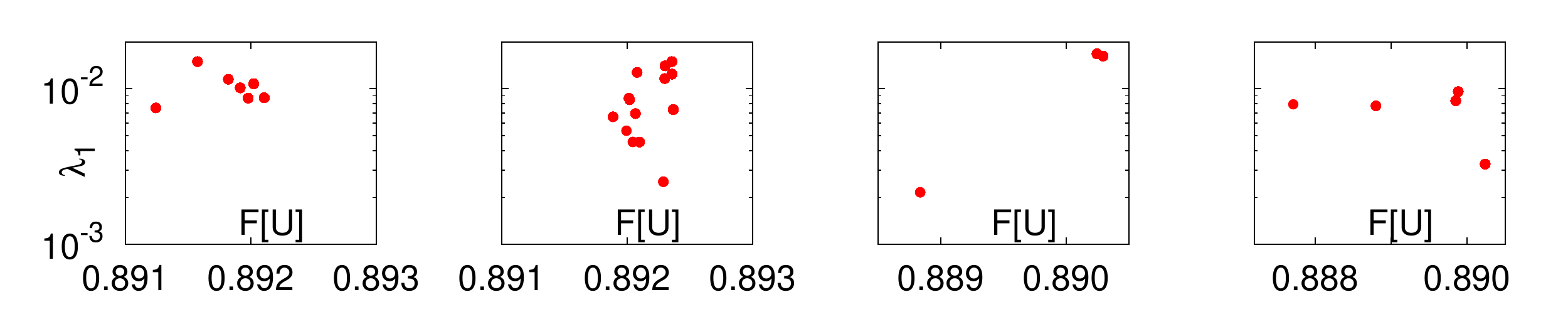}
 \caption{Smallest eigenvalue vs.\ functional value for 1000 copies from 4 arbitrary configurations of the $16^4$ 
lattices A1, A2, A3 (from top to bottom). The number of distinct Gribov copies decreases with finer lattices.}
\label{fig:gribov:lambda_vs_gff}
\end{figure}

Since the first Gribov region and the FMR have a common boundary, one may 
wonder if the bc-approach, which can be seen as an approximated search for 
configurations in the FMR, and the lc-approach, as an approximated search for
configurations close to the Gribov horizon, 
eventually converge to the same configuration. However, from the Landau gauge data
\cite{Sternbeck:2012mf} there is no such indication. Also for our Coulomb 
gauge data there is no evidence that the bc- and lc-procedure may coincide. In  
Fig.~\ref{fig:gribov:lambda_vs_gff} we show scatterplots for four arbitrarily 
selected configurations of each of the small ($16^4$) lattices A1, A2, A3 from top to 
bottom. The data points are from 1,000 different gauge copies, but there are much 
fewer points as the same Gribov copy is often found multiple times. In fact, the 
number of distinct Gribov copies varies strongly between configurations, compare for 
instance the third and fourth configuration of the A1 lattice (top right). As 
expected the number of distinct Gribov copies decreases with finer lattice spacing.

\begin{figure}[phtb]
  \includegraphics[width=0.49\columnwidth]{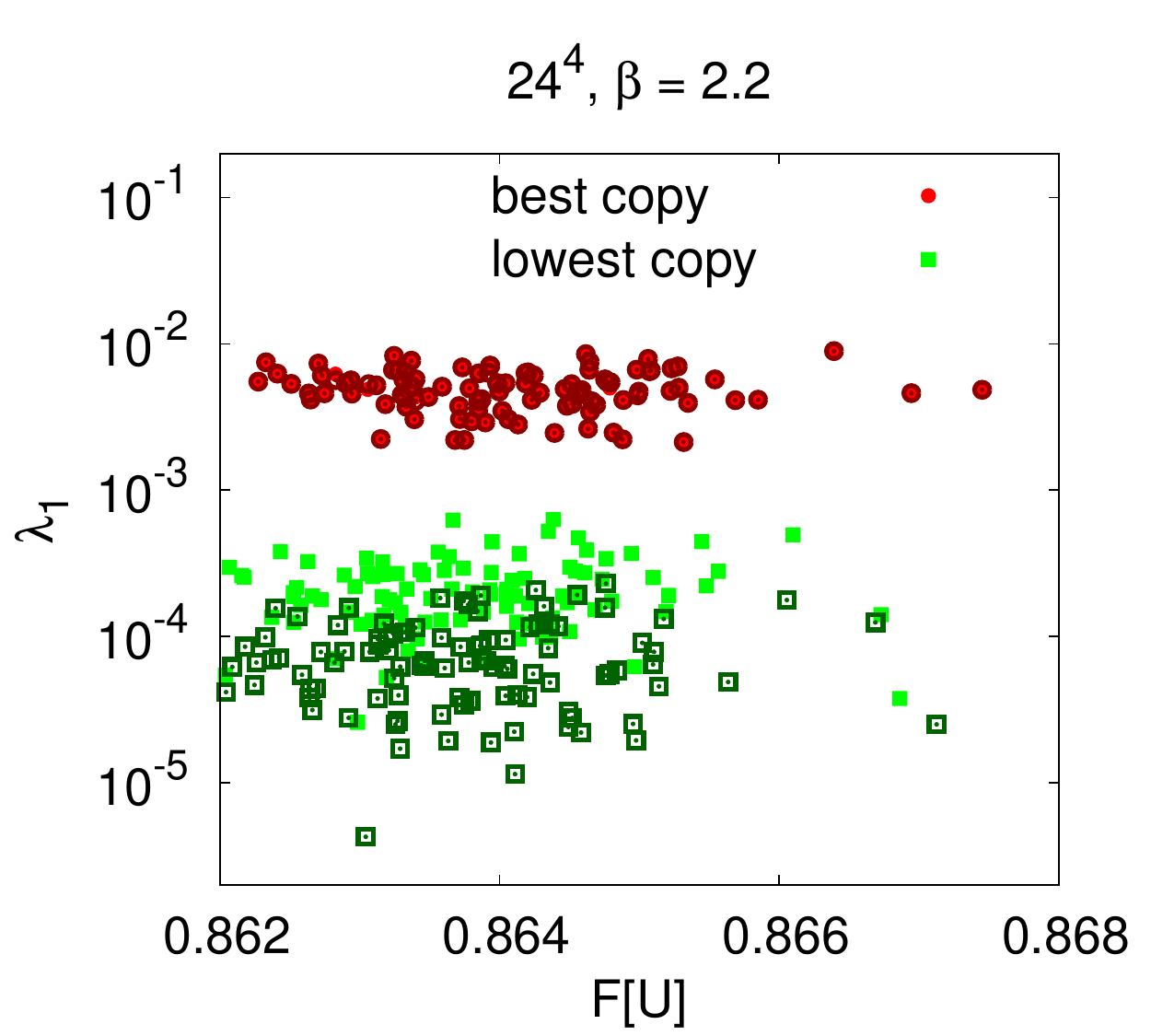}
  \includegraphics[width=0.49\columnwidth]{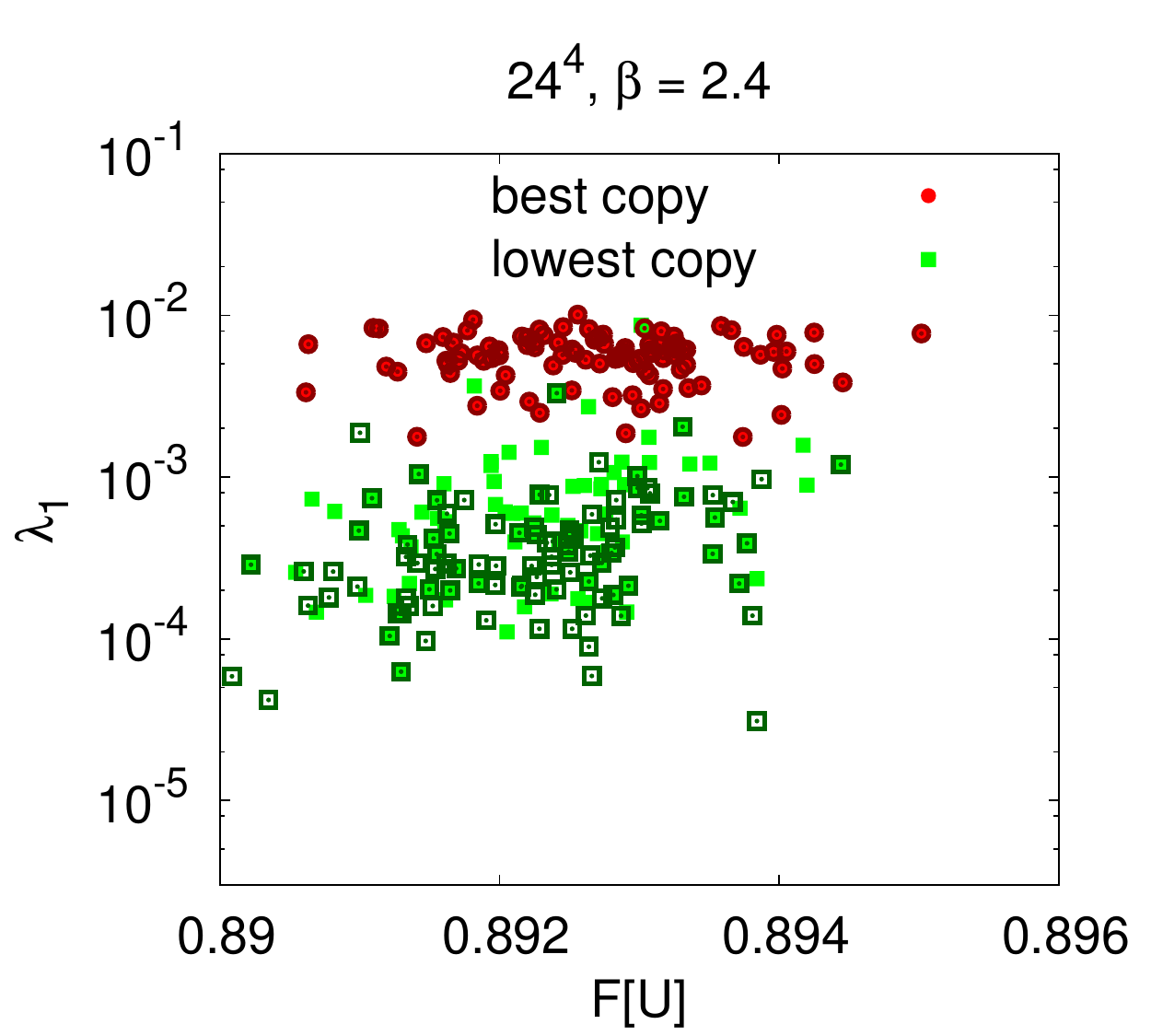}
 \caption{Best approximation of the FMR, i.e.\ the bc-approach, and the Gribov horizon, i.e.\ the lc-approach. Full 
light-colored symbols denote 1,000 trials; the empty, dark-colored symbols denote 10,000 gauge copies. Lattices: 
B1~(left), 
B3~(right). There is no configuration where bc = lc.}
 \label{fig:gribov:fmr_vs_GH}
\end{figure}

Another indication that bc and lc are different gauges is the result of 
Fig.~\ref{fig:gribov:fmr_vs_GH}. There we compare the best approximation of the FMR 
and the Gribov horizon for all 100 configurations of the $24^4$ sets B1 and B3 
after 1,000 and 10,000 gauge copies, respectively. Neither on the coarse lattice (B1) 
nor on the fine lattice (B3) could we find any configuration where the best-functional 
and the lowest-eigenvalue copy coincide. For the coarse lattice with 10,000 copies
the smallest eigenvalues are well separated, by at least an order of magnitude, in a first 
region with all the bc copies, and a second region with the lc copies.
While only very few (B1) or no configurations (B3) see a decrease of the lowest eigenvalue 
$\lambda_1$ in the bc case as we go from 1,000 to 10,000 copies, 
the lc data still sees a considerable reduction of $\lambda_1$.
A similar comparison was made for Landau gauge in Ref.~\cite{Maas:2015nva}. 
There, the authors used the value of the ghost propagator at the smallest non-zero
momentum
as an estimate of the smallness of the lowest FP eigenvalue. 
While they used a much larger ensemble of $\approx \mathcal{O}(10^3)$ configurations, 
they used much less gauge fixing repetitions $\approx \mathcal{O}(10)$. With this setup 
they found configurations that are close to \emph{both} the FMR and the Gribov horizon. 
However, it is clear that their setup (many configurations, small $N_r$) is specifically
biased towards finding such configurations, while our setup is biased in the 
opposite direction (fewer configurations, large $N_r$). For a detailed study of 
this effect, which is not our focus, we would have to significantly increase the statistics.

\begin{figure}[phtb]
  \includegraphics[width=0.49\columnwidth]{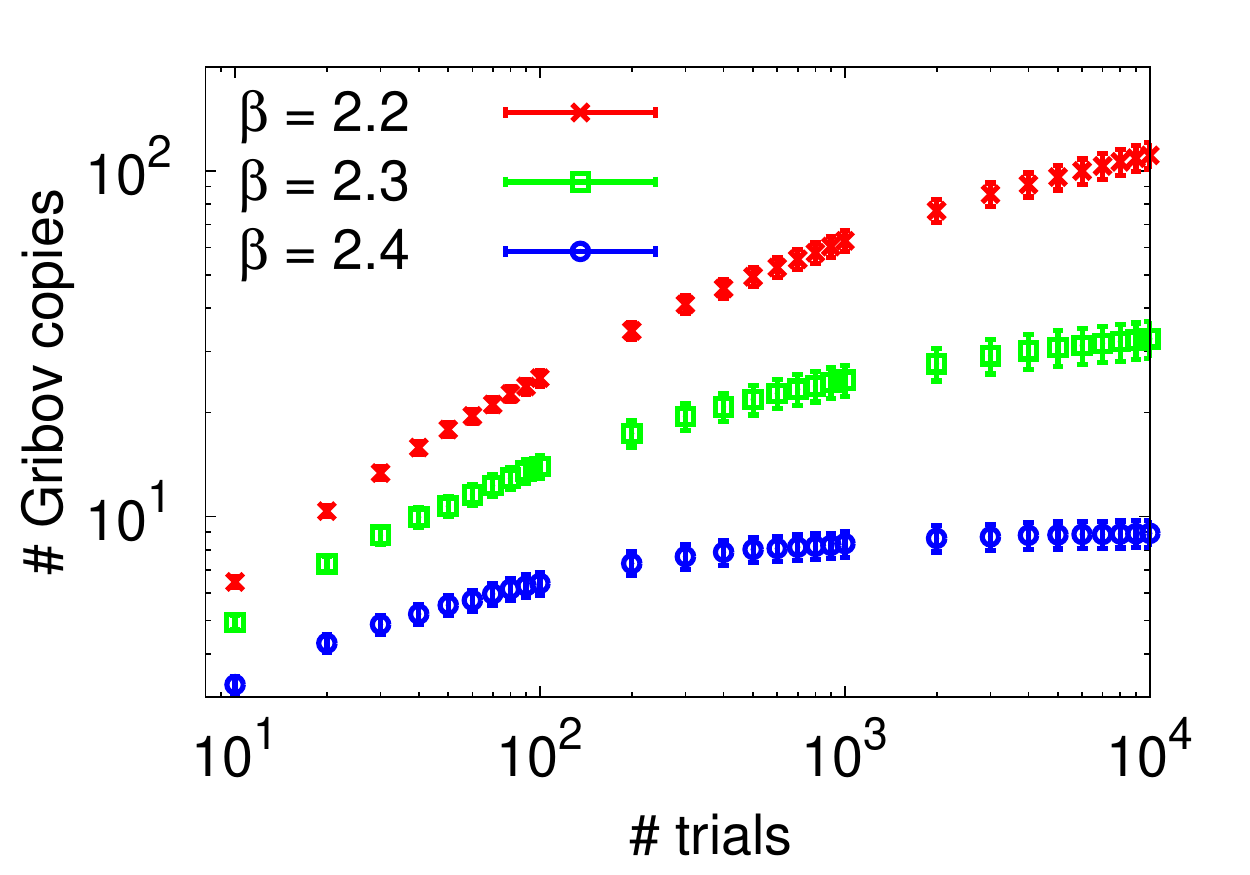}
  \includegraphics[width=0.49\columnwidth]{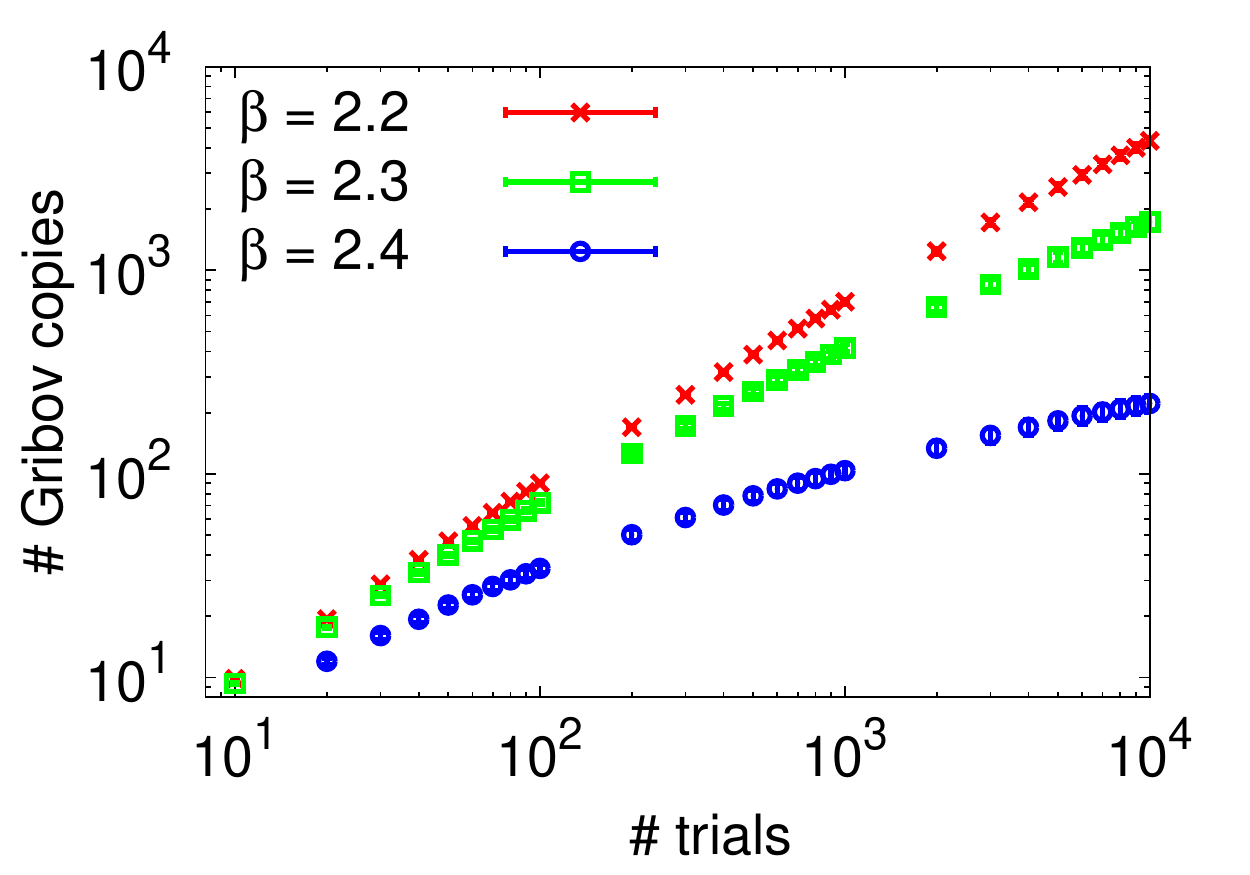}
 \caption{Number of distinct Gribov copies vs.\ the number of gauge copies for 
          the $16^4$ (left) and the $24^4$ (right) lattices. Only for the fines
          $16^4$ lattice a saturation is observed. The Gribov copy is 
          identified only by the value of the gauge fixing functional.}
\label{fig:gribov:number_gribovcopies}
\end{figure}

Finally, we try to estimate the number of Gribov copies in 
Fig.~\ref{fig:gribov:number_gribovcopies} by counting how many 
distinct Gribov copies we are able to find for a given number of g.f.~attempts. 
For this study we use only the functional value to identify the Gribov copy, 
since we do not have the smallest eigenvalue available for all copies (due to the 
threshold strategy described above). In general, an unambiguous identification
of a Gribov copy would require identical values for \emph{all} gauge-dependent 
quantities; the use of only a single quantity (the g.f.~functional) may 
therefore erroneously take distinct copies as identical, i.e.~the procedure 
is biased towards finding too many identical and too few distinct copies.\footnote{Additionally, 
the authors of Ref.~\cite{Maas:2015nva} found that there are gauge copies with same 
functional value but different value for the ghost propagator at smallest 
non-zero momentum.}. 
An unambiguous estimate would furthermore require that each Gribov copy is found with equal probability;
however very likely there are local maxima which are easier to locate by the algorithm. 
This effect will lead to an underestimation of the number of distinct Gribov copies. 
Thus the result in Fig.~\ref{fig:gribov:number_gribovcopies} 
has to be treated very carefully. More comprehensive studies of the
number of Gribov copies in lattice gauge theory can be found for example in 
Refs.~\cite{Hughes:2012hg,Mehta:2014jla}.
 
Since the number of Gribov copies varies considerably between different 
configurations, the error bars are rather large. Only on the smallest and 
finest lattice a saturation of the number of Gribov copies is observed within 
10,000 g.f.~attempts. The main conclusion we can draw from 
Fig.~\ref{fig:gribov:number_gribovcopies} is that we are far from having 
explored the whole Gribov region, which would be essential if the 
absolute lowest-eigenvalue copy still differs substantially from the 
lowest-eigenvalue copy in our limited search space.

\section{Results}
\label{sec:gribov:results}

While there is no compelling reason for the lc-approach to have a 
large effect on the gluon propagator, we expect a clear impact on the ghost 
propagator, given its spectral representation
\begin{equation}
 G(\vec p) = \sum_{n} \frac{\phi_n(\vec{p})\phi_n(\vec{-p})}{\lambda_n},
\label{eq_spectral}
\end{equation}
where $\lambda_n$ are the eigenvalues and $\phi_n(\vec p )$ the momentum space 
eigenfunctions of the \FP\ operator. As for the Coulomb potential, one also 
expects a large effect from the lc-strategy. Let us discuss them case by case.

\subsection{Gluon propagator}

\begin{figure}[hptb]
\center
  \includegraphics[width=0.8\columnwidth]{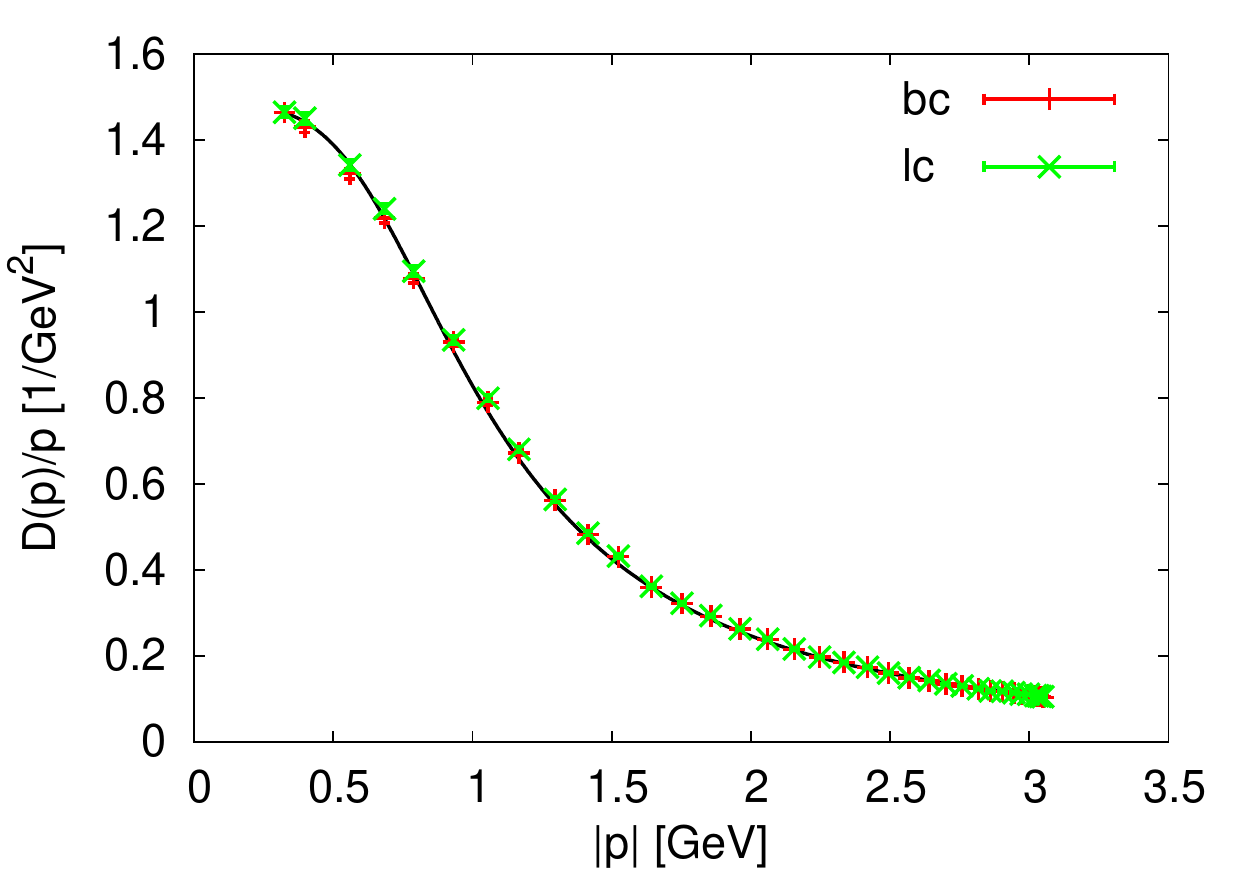}
 \caption{The gluon propagator for the B1 lattice with the bc and the 
          lc-approach from 1000 trials. The solid line is a fit to the Gribov 
          formula \cite{Burgio:2008jr,Burgio:2009xp}. The choice of Gribov copy
          apparently makes no visible difference.}
 \label{fig:gribov:gluonprop}
\end{figure}

In Landau gauge a small Gribov copy dependence was observed for the gluon 
propagator on a large $54^4$ lattice in the deep IR \cite{Sternbeck:2012mf}. 
With our lattice setup we are not able to reach that far in the IR and do not 
see a significant effect on the Coulomb gauge gluon propagator $D(\vec{p})$ as defined in 
Refs.~\cite{Burgio:2008jr,Burgio:2009xp}; see 
Fig.~\ref{fig:gribov:gluonprop}, where we plot $D(\vec{p})/|\vec{p}|$ to
underline its IR-behavior.\footnote{The expert reader will notice a strong similarity
between the IR-behavior of the gluon (and to a lesser extent ghost)
propagators in Coulomb and Landau gauge. This has been extensively discussed in
Ref. \cite{Burgio:2009xp,Burgio:2012bk} and can be simply ascribed to the
presence of common IR (Gribov) mass-scale in both cases.}
Since the accurate calculation of $D(\vec{p})$ requires the technique illustrated 
in Refs.~\cite{Burgio:2008jr,Burgio:2009xp}, Coulomb gauge needs to be fixed for 
\emph{all} timeslices. This limits the number of g.f.~repetitions as compared 
to the study of \FP-operator dependent quantities in the following sections,
which can all be evaluated on a single time slice. 

\subsection{Ghost propagator}
\begin{figure}[phtb]
  \includegraphics[width=0.49\columnwidth]{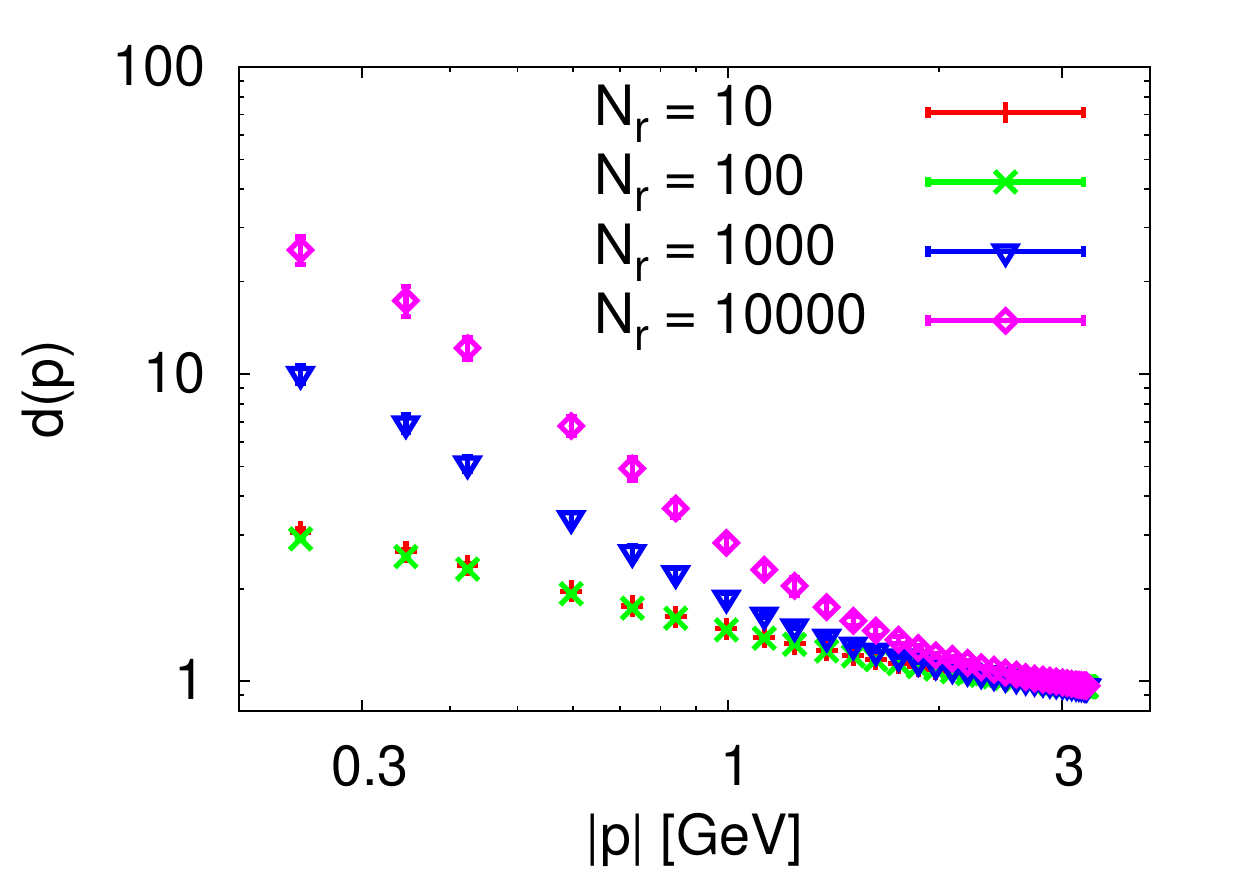}
  \includegraphics[width=0.49\columnwidth]{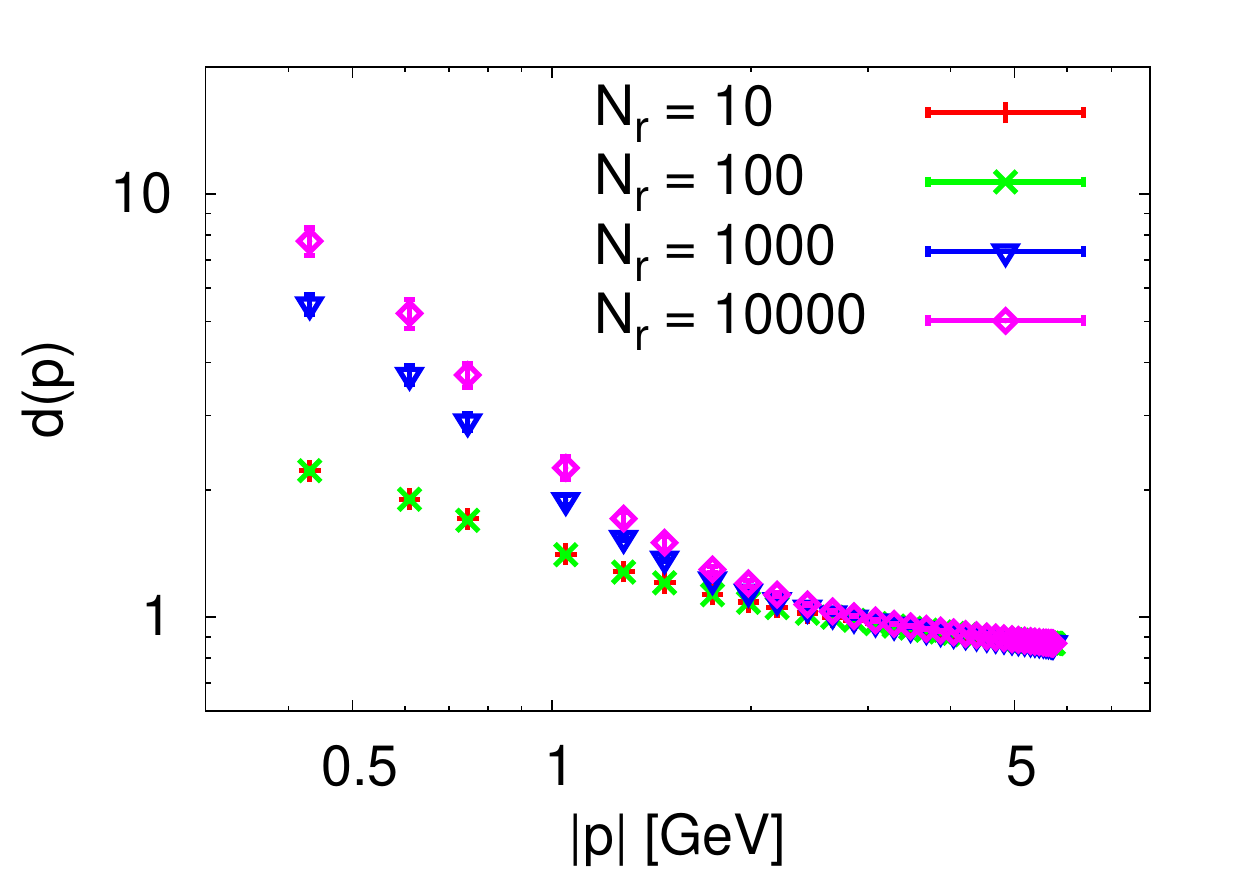}
 \caption{The ghost form factor after gauge fixing to the lowest-eigenvalue
          copy with increasing number of trials from 10 to 10,000 on $24^4$ 
          lattices at $\beta = 2.2$ (B1, l.h.s.) and $\beta = 2.4$ 
          (B3, r.h.s.).}
          \label{fig:gribov:ghost:lc:n24t24_allcopies}
\end{figure}

As expected from Eq.~\eqref{eq_spectral}, the Gribov copy effect (i.e.~the different
g.f.~prescriptions of picking Gribov copies) has a huge impact on the ghost propagator
as defined in Refs.~\cite{Burgio:2012bk}. 
In Fig.~\ref{fig:gribov:ghost:lc:n24t24_allcopies} we see that for the $24^4$ 
lattice the ghost form factor is drastically enhanced in the IR as the 
number of repetitions of the lc-strategy increases.  

For both the coarse and the fine lattice the effect first becomes 
visible upon reaching about 100 repetitions. From this point on, 
the form factor is clearly increased when going from 100 to 1,000 copies, 
while the further increase between 1,000 and 10,000 copies is less 
pronounced. This may be taken as a hint towards an eventual convergence, 
although no saturation of the effect can be really observed within our 
available data.  The huge difference in the IR is mainly due to a sharper 
bending in the region between 1 and 3 GeV. 

It should be noted that the data for different $\beta$ have been presented 
in different plots on purpose: the ghost form factor is known to suffer from 
scaling violations on isotropic lattices \cite{Burgio:2012bk}, so that data 
points for different $\beta$ do not fall on top of each other over the whole 
momentum range (after multiplicative renormalization). Moreover, since the curves 
in the lc-approach curves have not yet converged, the data from different $\beta$ 
cannot be compared, as the quality of the lc-gauge fixing for given $N_r$ most 
likely depends on the coupling $\beta$.

\begin{figure}[phtb]
  \includegraphics[width=0.49\columnwidth]{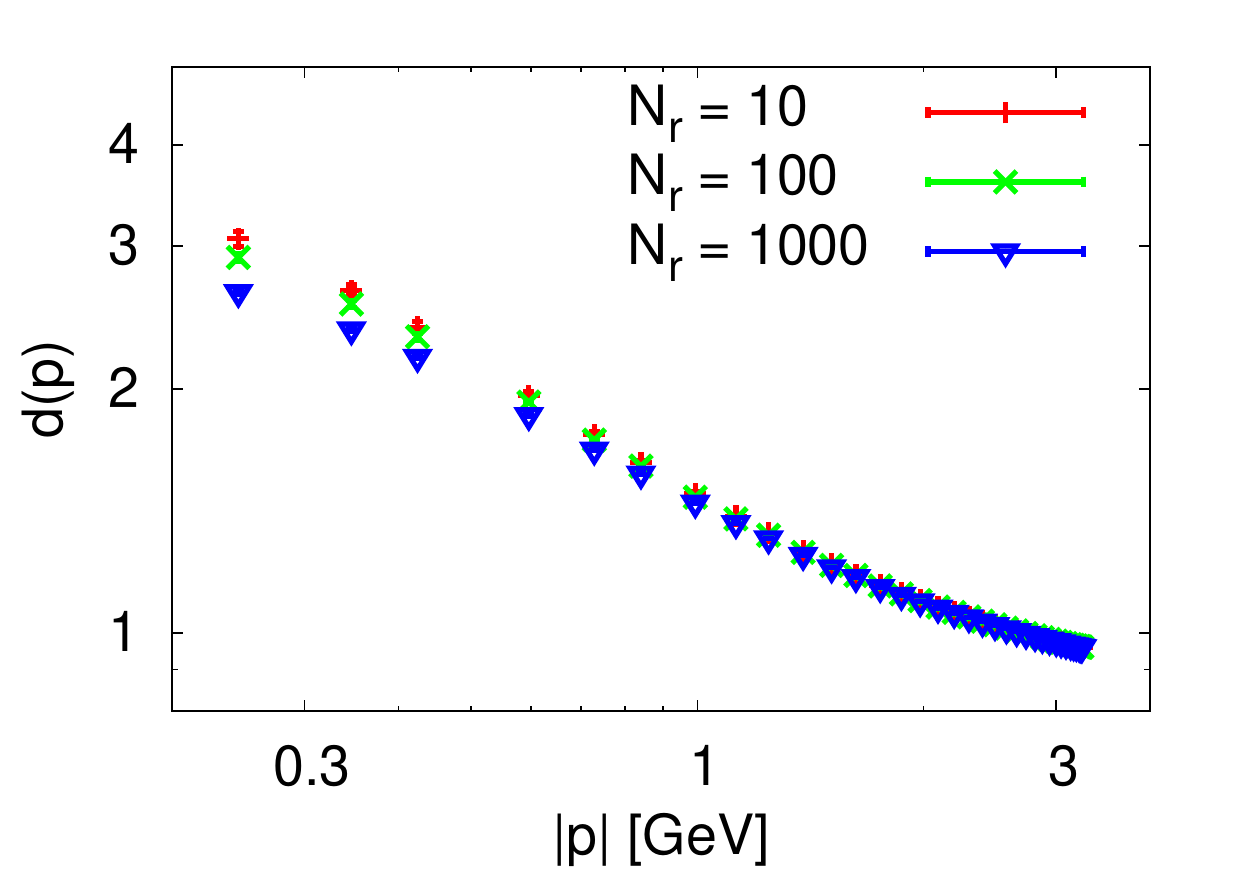}
  \includegraphics[width=0.49\columnwidth]{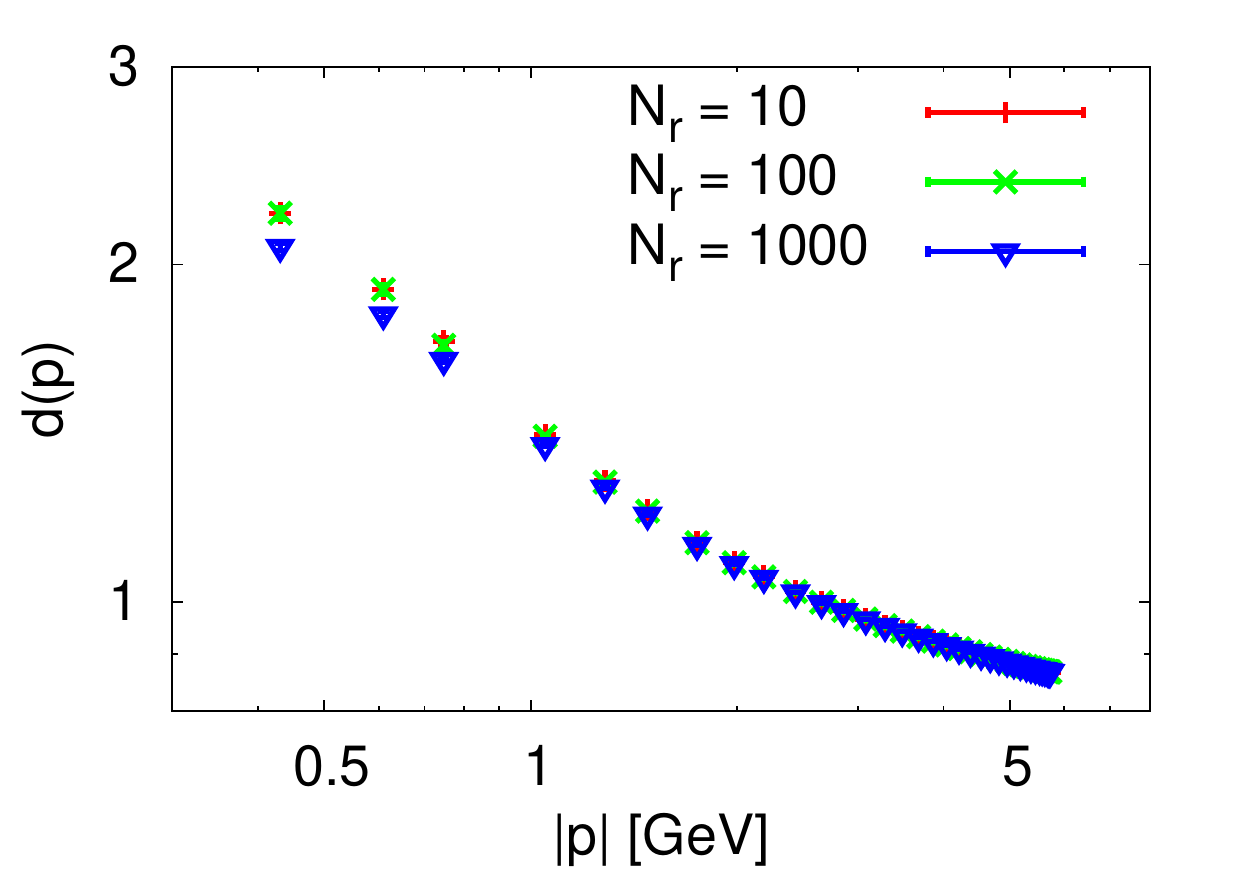}
 \caption{The ghost form factor after gauge fixing to the best-functional copy 
          with 
          increasing number of trials from 10 to 1,000 on $24^4$ lattices at 
          $\beta = 2.2$ (B1, l.h.s.) and $\beta = 2.4$ (B3, r.h.s.). The data 
          points for 10,000 copies are omitted since no better copy is found, 
          compare Fig.~\ref{fig:gribov:fmr_vs_GH}.}
 \label{fig:gribov:ghost:bc:n24t24_allcopies}
\end{figure}

\begin{figure}[phtb]
\center
  \includegraphics[width=0.8\columnwidth]{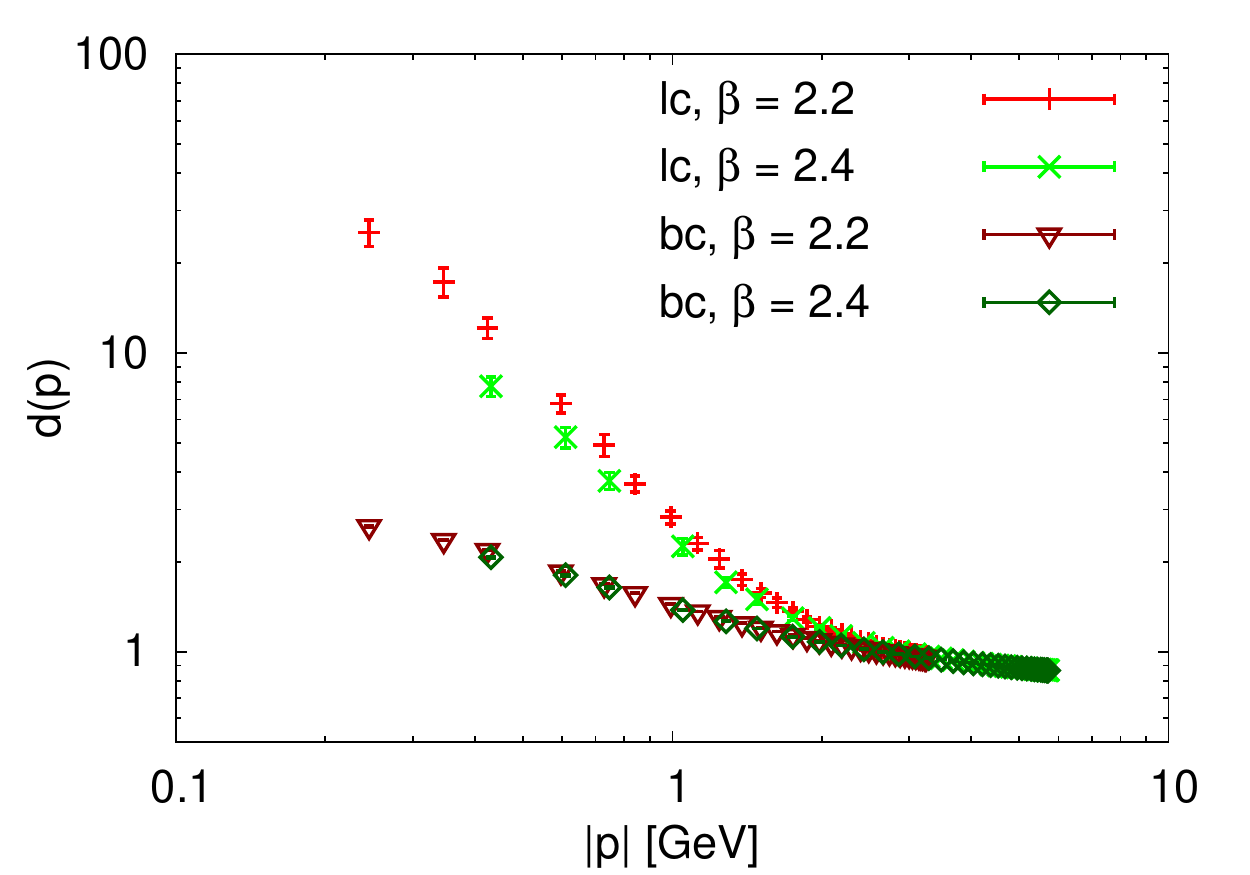}
 \caption{The ghost form factor from the lattices B1 and B3 after 10,000 copies 
          of bc and lc-strategy.}
 \label{fig:gribov:ghost:n24t24lcbc}
\end{figure}

In Fig.~\ref{fig:gribov:ghost:bc:n24t24_allcopies} we compare the ghost form 
factor within the bc-approach for the same lattices. 
First of all, the effect of taking more g.f.~repetitions is much less pronounced 
as compared to the lc-approach results in Fig.~\ref{fig:gribov:ghost:lc:n24t24_allcopies}. 
Secondly, the effect goes in the opposite direction: while the ghost form factor for the 
lc-approach was enhanced in the IR, the IR  form factor in the bc-approach becomes slightly 
suppressed as the number of g.f.~attempts is increased. The (small) effect is negligible between
10 and 100 repetitions, but becomes somewhat more pronounced in the region between 100  and 1,000  
repetitions. In Fig.~\ref{fig:gribov:ghost:n24t24lcbc} we compare the results for 
lc and bc-approach with 10,000 copies, our best values at this lattice size,
where we have renormalized the form factor to
\begin{equation}
 d(p=3 \,\giga \electronvolt) = 1\,.
\end{equation}
The bc-approach data at different $\beta$ fit quite well on top of each other, 
especially when considering the scaling violations~\cite{Burgio:2012bk}. 
Compared to the lc-strategy, the error bars for the bc-strategy are much smaller.

{To extract the IR-exponent of $d(p)$ we have fitted the data at different 
$N_r$ for the D1 ensembles in Tab.~\ref{tab:gribov:configs}. 
Since their UV-tail is not extended enough to extract reliably any UV-logarithmic
exponent, we used the simplest function interpolating between a power-law
in the IR and a constant in the UV:
\begin{equation}
 d(p) = p^{-\kappa} \frac{P_{n-1}(p)+a\, p^{n+\kappa}}{R_{n-1}(p)+p^n}
\label{eq:fit}
\end{equation}
where $P_{n-1}$ and $R_{n-1}$ are polynomials of 
degree $n-1$ and the denominator is constrained 
not to have any real poles (see e.g. Ref~\cite{Burgio:2012bk}). $n=2$
gives already a good enough fit and
the results are given in Fig.~\ref{fig:gribov:ghost:ani_22509} (continuous lines).
\begin{figure}[hptb]
\center
  \includegraphics[width=0.8\columnwidth]{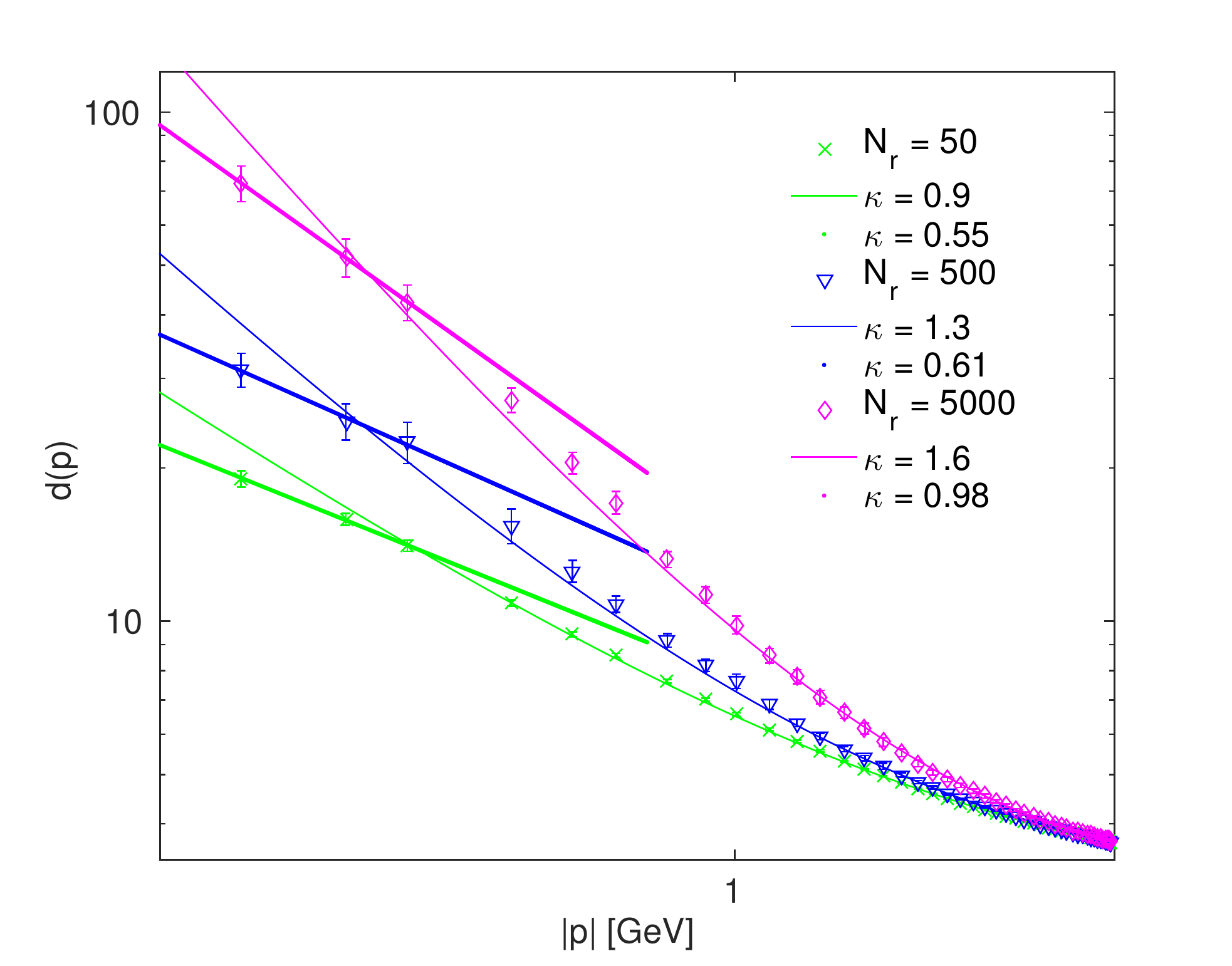}
 \caption{Fits of $d(p)$ to an IR-power law for the D1 lattices. The continuous lines are
 the fits to Eq.~\eqref{eq:fit}, the dotted lines just fit the last three points to
 a power law. The respective $\kappa$ values are given in the legend.}
 \label{fig:gribov:ghost:ani_22509}
\end{figure}
While in the bc-strategy we consistently found $\kappa ~ \simeq
0.5$ (see Refs.~\cite{Burgio:2012bk,Burgio:2013naa,Burgio:2015hsa}), for the lc-strategy
the exponent reaches $\kappa \simeq 0.9$ already for $N_r = 50$ 
and keeps on growing as $N_r$ increases, reaching
$\kappa \simeq 1.6$ for  $N_r = 5,000$, with no apparent saturation; 
the values of $\chi^2$/d.o.f. range between 0.9 and 1.4. The fits seem however to miss
the underlying behavior in the lowest IR region:
indeed, the good $\chi^2$/d.o.f. values come 
from the $p>1$ GeV data, while below these the curves clearly 
overestimate the IR behavior; changing $n$ does not improve
the situation. We have also tried to directly fit the last points to a power-law,
neglecting any sub-leading behavior: $d(p) = A \, p^{-\kappa}$. The results are also shown in 
Fig.~\ref{fig:gribov:ghost:ani_22509}. Although for $N_r = 5000$ we obtain
for $\kappa$ a value close to the continuum predictions, the evident lack of saturation
still means that by increasing $N_r$ we would probably overshoot $\kappa = 1$ again. Moreover,
the low momentum data are known to be effected by large IR-cut-off effects: only
simulations at higher volumes could eventually deliver reliable results.

All in all, the lack of saturation in the data will pose a challenge to any
fitting strategy, even if a theoretically sound  Ansatz for $d(p)$ over the whole
momentum range, going beyond 
a simple power law, could be found. 
We will see in the next section that such lack of saturation is an even bigger
problem for the calculation of the Coulomb potential.}

\subsection{Coulomb potential}

\begin{figure}[phtb]
  \includegraphics[width=0.49\columnwidth]{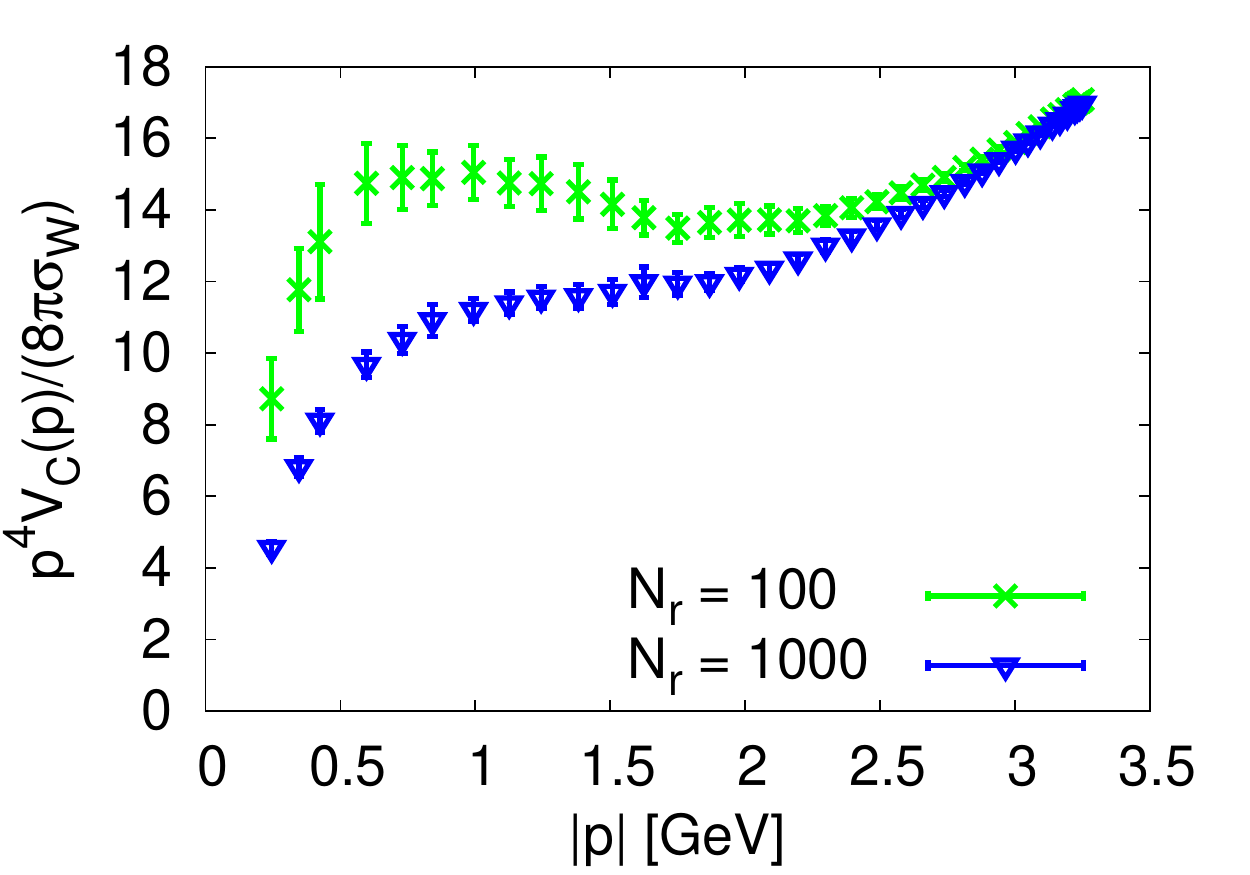}
  \includegraphics[width=0.49\columnwidth]{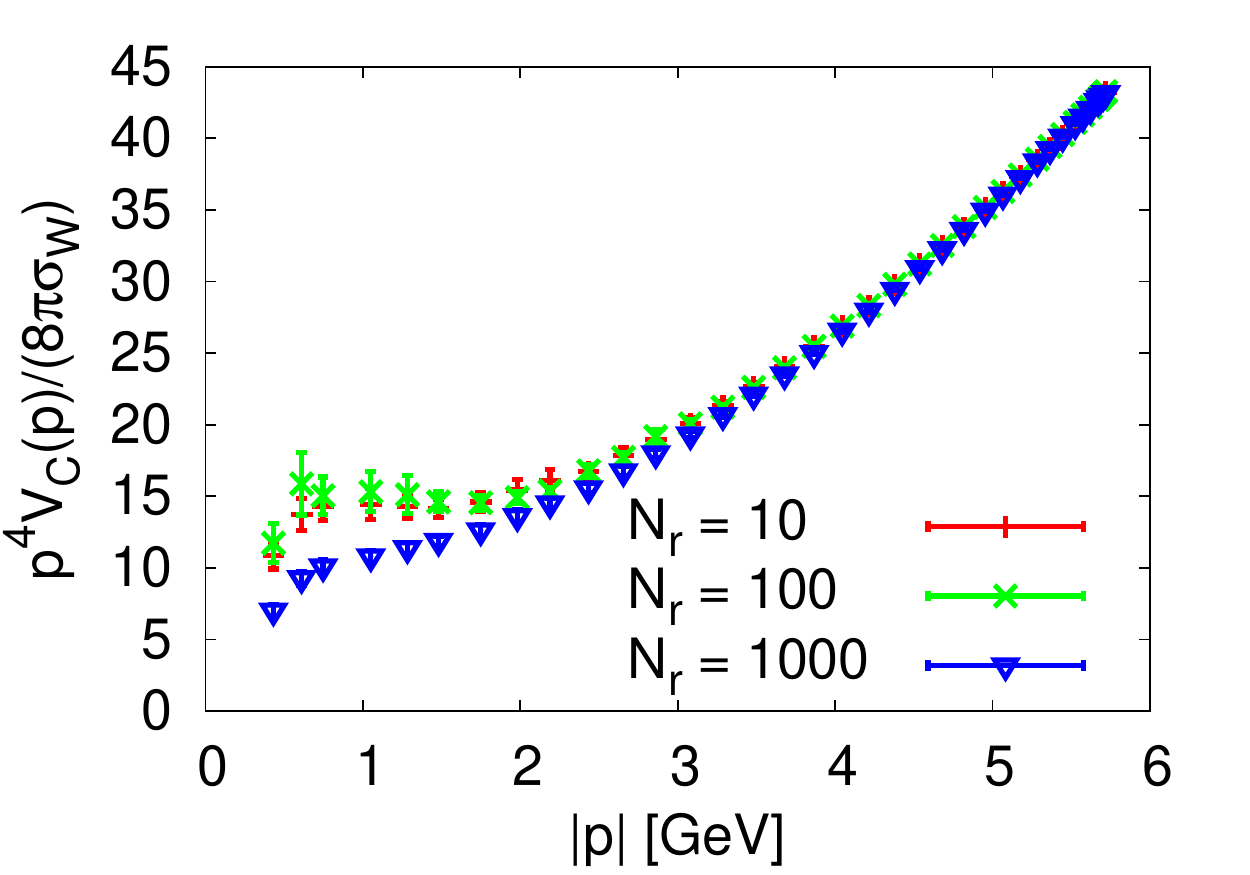}
 \caption{Coulomb potential in the bc-approach as a function of
 $N_r$. The data with $N_r=10,000$ are omitted, since they show no difference as compared
 to $N_r=1,000$, see Fig.~\ref{fig:gribov:fmr_vs_GH}. 
 }
 \label{fig:gribov:coulpot:n24t24_allcopies}
\end{figure}

The most important quantity for Coulomb gauge confinement is the Coulomb 
potential, since it provides direct access to the Coulomb string tension; 
this quantity can be computed from the momentum space Coulomb kernel
\cite{Burgio:2012bk}:
\begin{equation}
\label{eq:coulombpotential}
V_C(\vec{p}) = g^2\, \mathrm{tr}\left\langle \left( -\hat{\vec{D}} \cdot \nabla 
\right)^{-1} \left(-\nabla^2\right)\left( -\hat{\vec{D}} \cdot \nabla \right)^{-1}
\right\rangle\,.
\end{equation}
A linearly rising potential for large distances corresponds to a momentum 
space potential diverging like $1/p^4$ in the IR. 
Thus, it is convenient to plot the potential such that its intercept with 
the y-axis yields the Coulomb string tension $\sigma_C$ in units of the 
physical (Wilson) string tension $\sigma$, 
\begin{equation}
 \frac{p^4V_C(p)}{8\pi\sigma} \xrightarrow{p \rightarrow 0} 
\frac{\sigma_C}{\sigma}\,.
\label{eq_CP}
\end{equation}
In Fig.~\ref{fig:gribov:coulpot:n24t24_allcopies} the ratio eq.~\eqref{eq_CP} 
is shown, within the bc-approach, for the same configurations used in 
Fig.~\ref{fig:gribov:ghost:bc:n24t24_allcopies} for the ghost propagator.\footnote{In 
the $\beta=2.2$ plot on the left hand side the data 
for $N_r = 10$ is omitted since it contained a configuration with a small 
eigenvalue leading to very big error-bars. We 
will discuss the issue in more  detail below.}
In earlier studies of the Coulomb potential a bump in $p^4V_C(p)$ 
was observed at around 0.5 to 1 $\giga \electronvolt$, affecting direct estimates of the
intercept on the vertical axis with large uncertainties 
\cite{Nakagawa:2008zza,Voigt:2008rr,Greensite:2009eb,Burgio:2012bk,Burgio:2013naa,Burgio:2015hsa}. 
As Fig.~\ref{fig:gribov:coulpot:n24t24_allcopies} shows,
this bump does indeed vanish as the number of gauge copies is increased; at the same time 
the statistical precision on the MC-data strongly improve. 
One might be tempted to assume that one is actually approaching the 
absolute maximum of the gauge fixing functional  as the number of gauge 
copies is increased, and the ensemble eventually samples a FMR free of 
Gribov copies.\footnote{We had put forward such hypothesis in 
Ref.~\cite{Vogt:2013jha}, although in a slightly different context.}
If this was the case, however, one should expect that the Coulomb potential
from the alternative lc-approach  should yield the same (or a similar) result,
as the Gribov-Zwanziger entropic argument in the 
thermodynamic limit states that the partition function is dominated
by configurations lying on the common boundary of the FMR and 
the first Gribov region. For such configurations, the bc and lc approach -- 
once they converged -- would give identical results. 

\begin{figure}[phtb]
\center
  \includegraphics[width=0.8\columnwidth]{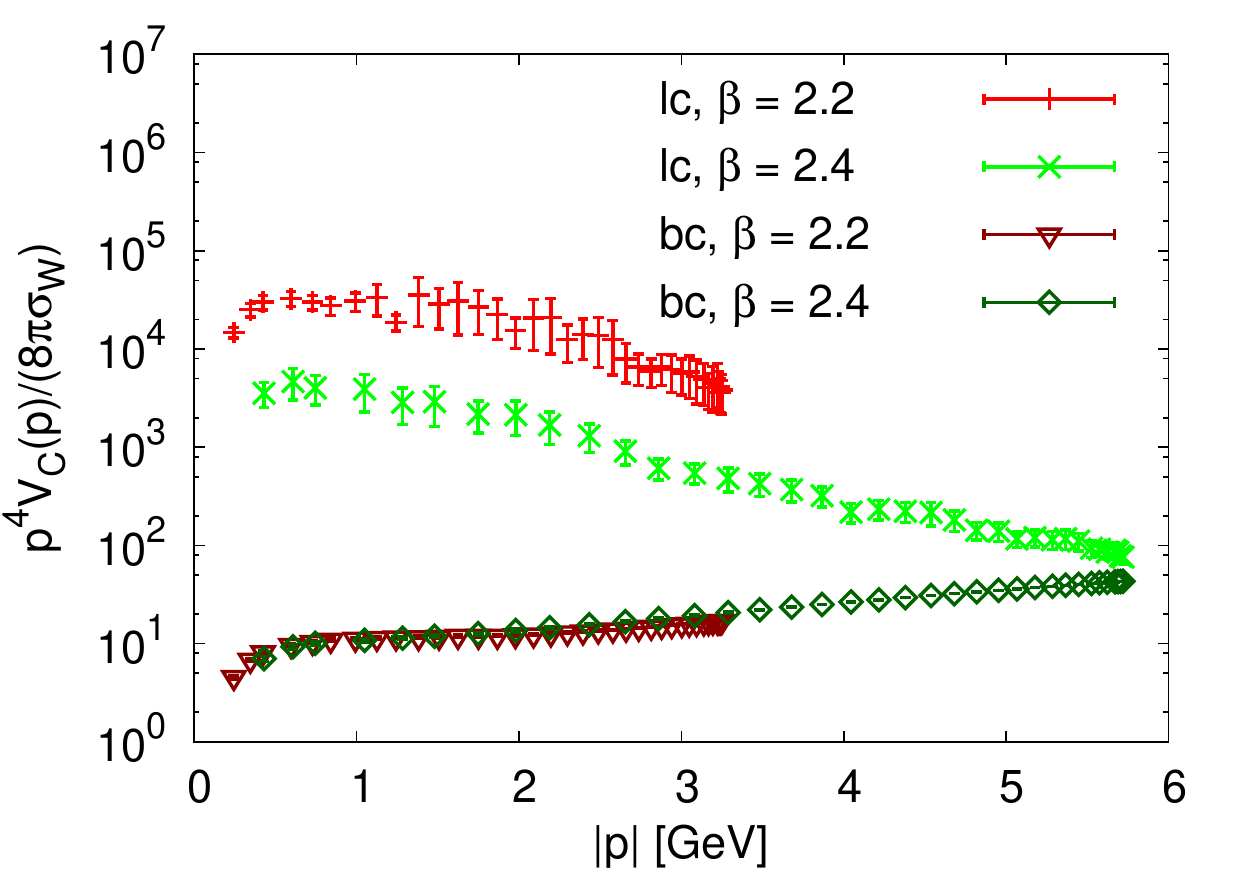}
 \caption{The Coulomb potential from the lattices B1 and B3 after 10,000 copies.}
 \label{fig:gribov:coulpot:n24t24lcbc}
\end{figure}

Unfortunately, the lc-result for the Coulomb potential does not corroborate such
a conjecture. In 
Fig.~\ref{fig:gribov:coulpot:n24t24lcbc} the data for the bc- and the lc-approach 
are compared for the B1 and B3 
lattices. 
While for the ghost 
propagator the different gauge fixing strategies provided a nice overlap in 
the UV regime (see Fig.~\ref{fig:gribov:ghost:n24t24lcbc}), the Coulomb potential, over 
the whole momentum range, is increased by \emph{several orders of magnitude!}  
The same happens for all lattices that we investigated.
Since this result is quite surprising, we have repeated the calculation with a 
different solver. We have usually adopted a conjugate gradient algorithm with 
Laplace pre-conditioning. To ensure the validity of our solver for 
exceptional configurations\footnote{The lc-strategy generally attempts to make 
the \FP\-operator ill-conditioned, but for some configurations with a very  small
eigenvalue $\lambda_1$, it becomes nearly  singular and its precise inversion in the 
Coulomb potential is numerically challenging.} we compared 
the results of our conjugate gradient to a publicly available C++ 
implementation \cite{Villa2012} of the MINRES algorithm \cite{Paige:1975:SSI}. 
Both algorithms yield the same solution up to numerical precision.

\begin{figure}[phtb]
\center
  \includegraphics[width=0.8\columnwidth]{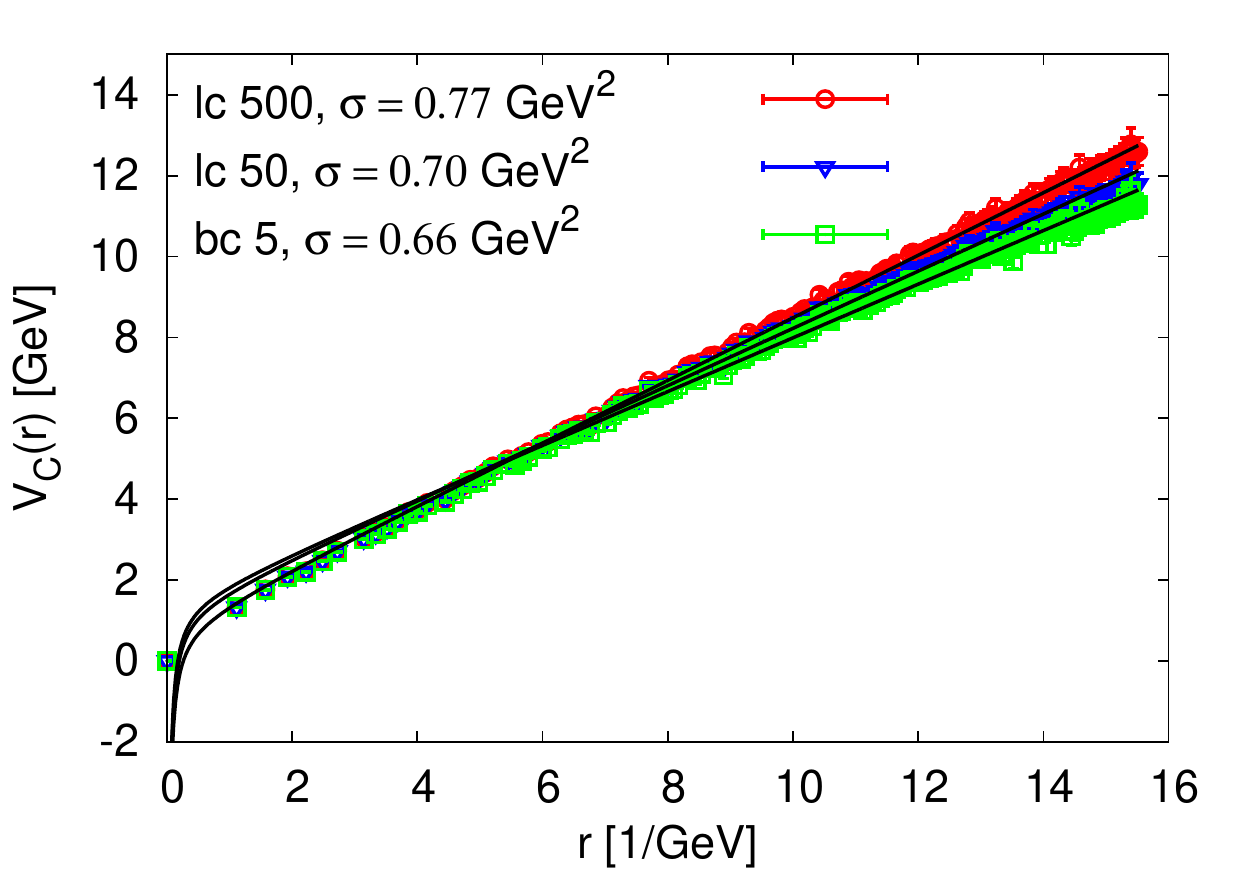}
 \caption{Coulomb potential in position space from the 
          $\left<U_0\,U_0^\dag\right>$ correlator in eq.~(\ref{eq:correlator})(D1 lattice). 
          The bc configurations are preconditioned with simulated annealing.}
 \label{fig:gribov:u0u0_b22509}
\end{figure}

Interestingly, while the Coulomb potential in the lc-approach computed from the kernel 
Eq.~\eqref{eq:coulombpotential} apparently yields physically non-sensible
results, the alternative definition proposed in Refs.~\cite{Marinari:1992kh,Greensite:2003xf},
which is based on short Polyakov lines $P_t$ of length $t$  and the correlator of 
temporal links $U_0$, 
\begin{align}
\label{eq:correlator}
 a V_C(|\vec{x}-\vec{y}|) &= -\lim_{t\rightarrow 0} \frac{d}{dt} \log \left\langle 
\tr\,{P_t(\vec{x})P_t^\dagger(\vec{y})}\right\rangle\nonumber\\[2mm]
  &= -\log \left\langle \tr\,{U_0(\vec{x})U_0^\dag(\vec{y})}\right\rangle\,,
\end{align}
seems to work in all cases, cf.~Fig.~\ref{fig:gribov:u0u0_b22509}.
As in the case of the gluon propagator, the effect 
of choosing different g.f.~strategies and selection of Gribov copies is quite modest. 
To extract the Coulomb string tension we fitted $V_C$ from 
Eq.~\eqref{eq:correlator} to
\begin{equation}
 V_C(r) = \frac{\alpha}{r}+\sigma_C\, r + \text{const.}\,,
 \end{equation}
where the L\"uscher-term $\alpha = -\frac{\pi}{12}$ is kept fixed.
In the range $[6/a,14/a]$ we find a Coulomb string tension varying between 
$(0.66 \,\giga \electronvolt)^2$ (bc 5) 
and $(0.77 \,\giga \electronvolt)^2$ (lc 500), with $\chi^2$/d.o.f. between $0.58$ (lc 500)
and $0.95$ (bc 5).

\section{Conclusions}

In this paper we have studied the effect of fixing, on the lattice, 
the Coulomb gauge to the copy closest to the Gibov horizon,
i.e.~the copy with the smallest non-vanishing eigenvalue of the 
\FP\ operator (\emph{lc-approach}). This prescription  
{\it de facto} implements the proposal of Ref.~\cite{Cooper:2015sza}. 
The main observation we made is that the size of the smallest eigenvalue
saturates very slowly, if at all, with the number of gauge-fixing 
attempts, see e.g.~Fig.~\ref{fig:gribov:fmr_vs_GH}. Of course
we are still far from exploring the whole Gribov region, as
Fig.~\ref{fig:gribov:number_gribovcopies} suggests; still
our result is somehow surprising: in light of the entropic argument
usually made within the Gribov-Zwanzier scenario, one would have 
intuitively expected that the small eigenvalue of the \FP\ operator 
should be bounded from below by some effective IR cutoff induced by the 
finite lattice volume; we however see no such saturation even after 
$N_r = 10,000$ gauge fixing repetitions.

The small eigenvalues heavily affect the IR behavior of the ghost propagator 
and, more importantly, the Coulomb potential extracted from the kernel in 
Eq.~\eqref{eq:coulombpotential}. {The first effect can be regarded as 
positive to some extent, as it moves the infrared behaviour of the ghost propagator 
towards the continuum prediction and thus reduces the violation of the sum 
rule. As with the size of the smallest eigenvalue, we do not see a saturation 
with the number of gauge copies, and the ghost exponent eventually tends to 
overshoot the continuum prediction. However, given the arbitrariness in the fits
used to extract the exponent, it is at least conceivable that the \emph{lc approach}
could indeed be made to agree with the continuum.}

{Much more severe is the 
second effect, on the Coulomb potential, which yields results that are 
at odds with physical expectations.}
The dramatic increase of the potential extends 
of the entire momentum range and also affects the Coulomb string tension, to 
the point that the results are physically non-sensible. As the effect on the
eigenvalues has not
yet saturated with $N_r = 10000$ gauge fixing repetitions, exploring the Gribov
region further by increasing $N_r$ should make things even worse. 

We can think of {several} possible interpretations of our result.
First, it could be that merely constraining the lowest eigenvalue is
insufficient to detect the physical relevant configurations. From the 
entropic argument one expects the partition function to be peaked on the common
boundaries of the first Gribov region and the fundamental modular region, 
i.e.~on configurations where the absolute maxima of Eq.~\eqref{eq:gribov:gff} become 
degenerate. The multiple flat directions allow for further refinements of the 
lc-prescription; for instance, a restriction to configurations where {\it at least} 
the two lowest eigenvalues are small {\it and} (nearly) degenerate could lead to 
the correct physics. Such an investigation is, although numerically demanding, in 
principle feasible and its implementation is currently under scrutiny.

A second possibility is that the Coulomb potential as calculated 
from Eq.~\eqref{eq:coulombpotential} involves the inverse of the 
ill-conditioned lattice \FP\ operator whose kernel may be sensitive 
to the exact lattice definition and (yet to be determined) discretization artifacts.
The lc-procedure would then bring this defect to the fore and amplify it, ultimately 
making the lattice definition impractical. The fact that the alternative definition
given in Eq.~\eqref{eq:correlator}, which requires no such operator inversion,  
always works well might indeed point in this direction. 
Also the fact that no saturation for the smallest eigenvalue could be found 
hints towards spurious artifacts in the low-lying spectrum of the lattice 
\FP\ operator. If this  issue could be resolved and a saturation could be found, 
it is also conceivable that a theoretically motivated \emph{Ansatz} for the fit to 
the data in Fig.~\ref{fig:gribov:ghost:ani_22509} might still bring the results 
in agreement with the continuum predictions, e.g.~the ghost exponent $\kappa=1$. 

Alternatively one could argue that, since the fundamental discretization 
of Yang-Mills theory is known to possess lattice artifacts which affect gauge invariant, 
topological observables \cite{Barresi:2004qa,Burgio:2006dc,Burgio:2006xj,Burgio:2014yna}, 
it {is} conceivable for them to also influence gauge dependent quantities.
In this case, it is the discretization of the model itself which would introduce spurious 
quasi-zero modes in the FP operator which subsequently affect all quantities that require 
its inversion (such as the ghost propagator or the Coulomb potential). By contrast, 
ordinary correlators that require no FP inversion are benign, cf.~Eq.~\eqref{eq:correlator}.
To test such a hypothesis one would, however, need to explore Coulomb gauge in algorithmically 
demanding alternative discretizations of the Yang-Mills action 
\cite{Barresi:2003jq,Barresi:2006gq,Burgio:2006dc,Burgio:2006xj}.

{Finally, it is also conceivable that the GZ scenario realised in the 
Hamiltonian approach does not describe the lattice results at all, and a refinement 
of the Coulomb Hamiltonian would be necessary, similarly to what was 
conjectured in Landau gauge \cite{Cucchieri:2011ig}. If such a refinement is to 
remain renormalizable, however, the additonal terms would dominantly affect the 
infrared regime, and hence the sum rule, but they could not explain for the dramatic 
increase of the Coulomb potential observed in the \emph{lc} approach over the 
\emph{entire} momentum range. More 
generally, the numerical investigations in this paper show that there is not one 
single consistent version of Coulomb gauge on the lattice, at least not within 
current computational limits,  and it is hence unclear in which way to extend 
the continuum GZ scenario.  At the moment, we have no strong evidence for an 
extension of the present continuum formulation, i.e. the standard Hamiltonian 
approach realising the GZ horizon condition remains our best continuum 
description so far.}

A by-product of our investigation was the systematic improvement of the 
search for the best gauge functional value (bc-approach) with a high 
number of g.f.~repetitions. While in this case gluon and ghost propagators 
(Figs.~\ref{fig:gribov:gluonprop} and \ref{fig:gribov:ghost:bc:n24t24_allcopies}) 
do not change as compared to previous investigations \cite{Burgio:2008jr,%
Burgio:2009xp,Burgio:2012ph,Burgio:2012bk,Burgio:2013naa,Burgio:2015hsa},
the Coulomb potential loses the ``bump" in the low momentum region found
in previous works, which allows for a much more reliable estimate of the Coulomb
string tension in this setup. For a true high-precision determination of $\sigma_C$, 
however, larger volumes and a systematical finite-size analysis would be required.  

\begin{acknowledgments}
This work was partially supported by the Deutsche Forschungsgemeinschaft under the 
contract DFG-Re 856/10-1.
H.V. wishes to thank the Evangelisches Studienwerk Villigst for financial 
support.
\end{acknowledgments}

\bibliographystyle{apsrev4-1}
\bibliography{thesis}

\end{document}